\documentclass[pre,preprint,showpacs,amsmath,amssymb]{revtex4}
\usepackage{bm}
\usepackage{graphicx}

\begin{document}

\title{\textbf{Hydrodynamic Modes for a Granular Gas from Kinetic Theory}}
\author{J. Javier Brey}
\affiliation{F\'{\i}sica Te\'{o}rica, Universidad de Sevilla,
Apartado de Correos 1065, E-41080, Sevilla, Spain}
\author{James W. Dufty}
\affiliation{Department of Physics, University of Florida,
Gainesville, FL 32611}
\date{\today }

\begin{abstract}
Small perturbations of the homogeneous cooling state (HCS) for a
low density granular gas are described by means of the linearized
Boltzmann equation. The spectrum of the generator for this
dynamics is shown to contain points corresponding to hydrodynamic
excitations. The corresponding eigenvectors and eigenvalues are
calculated to Navier-Stokes order and shown to agree with those
obtained by the Chapman-Enskog method. The conditions for the
hydrodynamic excitations to dominate all other excitations are
discussed.
\end{abstract}


\maketitle

\section{Introduction}
The use of hydrodynamic equations to describe granular fluids in
rapid flow has been in practice for many years \cite{Ha83}. The
justification for this fluid-like description and prediction of
the transport coefficients appearing in these equations has been
the focus of attention for some time as well
\cite{JyR85,LSJyCh84,GyS95}. In recent years, an accurate
derivation of Navier-Stokes order hydrodynamics has been given
from the Boltzmann equation for granular gases using an adaptation
of the Chapman-Enskog method for\ normal gases
\cite{BDKyS98,GyS95}. The expressions for the transport
coefficients as a function of the degree of inelasticity have been
confirmed by both Monte Carlo and molecular dynamics simulation
\cite{ByC01}. Successful application of these Navier-Stokes
equations to a number of different states also supports their
validity \cite{BRyC99,Go03,ByR04}. However, the context of the
hydrodynamic equations remains uncertain. What are the relevant
space and time scales? How much inelasticity can be described in
this way?

Such questions can be addressed for gases using the Boltzmann
kinetic equation to describe the complete dynamics for properties
of interest. Then it can be asked under what conditions do the
hydrodynamic excitations emerge as the dominant dynamics. The
analysis of this problem for normal gases is quite complete and
precise \cite{Mc89,Scharf}. The objective here is to initiate a
similar formulation of the  problem for granular gases
\cite{DyB03}. First, solutions to the Boltzmann equation are
considered for states that deviate from spatial homogeneity only
by small perturbations. The dynamics in this case is governed by
the \emph{linear} Boltzmann operator for spatially inhomogeneous
states. The spectrum of this operator determines all possible
excitations on all space and time scales, and for all degrees of
inelasticity. The first problem is to identify points in this
spectrum corresponding to hydrodynamics. This is one of the main
results reported here. Both the hydrodynamic eigenvalues and
eigenfunctions are calculated for long wavelength excitations
corresponding to Navier-Stokes order hydrodynamics. Their
agreement with the corresponding results from the Chapman-Enskog
method is established.

Next, the issue of conditions for the dominance of  the
hydrodynamic excitations, or modes, is considered. This entails
showing that the hydrodynamic eigenvalues are smaller than all
other parts of the spectrum, such that there is a long enough time
scale for the latter to decay relative to the hydrodynamic modes.
There are two new difficulties for granular gases. First, the
hydrodynamic eigenvalues cannot be made arbitrarily small since
they do not all scale with the inverse wavelength of the
perturbation, as  for normal gases. Second, there is a new
characteristic frequency, the cooling rate, in addition to the
collision frequency to set the scale of the spectrum. The cooling
rate can be relatively large or small depending on the
inelasticity of the gas particle collisions.

Analysis of the non-hydrodynamic spectrum of the Boltzmann
equation for a granular gas remains a difficult unsolved problem.
Instead, we consider here a simpler kinetic model \cite{BDyS99}
that retains the exact hydrodynamic spectrum and allows
identification of the entire spectrum as well. It is found that
the time scale for the hydrodynamic excitations is longer than
that for all  other excitations, for any degree of inelasticity. A
brief report of these results has been given in reference
\cite{DyB03}, with the details and elaboration given below.

The plan of the paper is as follows. In the next section, a short
summary of the Boltzmann equation is given, and the Navier-Stokes
equations derived from it by the Chapman-Enskog procedure are
recalled. More details are given in Appendix \ref{ap1}. Also, the
hydrodynamic equations are linearized and the hydrodynamic modes
to second order in the wavevector are identified.

In Sec. \ref{s3}, the structure of the linear Boltzmann equation
is discussed, emphasizing the relevance of the spectrum of the
operator generating the linear dynamics to determine the existence
and validity of hydrodynamics. The first of the above issues is
addressed in Sec.\ \ref{s4}, where the kinetic theory hydrodynamic
modes are identified in the long wavelength  limit. A technical
point associated with the non Hermitian character of the operator
is discussed in Appendix \ref{ap2}. The results are extended to
Navier-stokes order in Sec. \ref{s5}, and the obtained expressions
are shown to agree with those derived in Sec. \ref{s2} from the
hydrodynamic equations. Details of the calculations are given in
Appendix \ref{ap3}. The possibility of a description in terms of
the hydrodynamic modes is studied in Sec.\ \ref{s6}, by means of a
model kinetic equation. It is established that there are length
and time scales on which only the excitations associated to the
hydrodynamic modes persist. The mathematical details are given in
Appendix \ref{ap4}. Finally, the last section contains a short
summary of the main points and conclusions in the paper.

\section{Boltzmann Equation and Hydrodynamic Modes}
\label{s2} In all of the following, the simplest model of a
granular fluid is considered: a low density gas of smooth and
inelastic hard spheres ($d=3$) or disks ($d=2$) of mass $m$ and
diameter $\sigma $. The binary collision rule given below is
characterized by a constant coefficient of restitution $\alpha$,
defined  in the interval $0<\alpha \leq 1$, and measuring the loss
of energy in each collision. At sufficiently low density, the
distribution function $f({\bm r},{\bm v},t)$ is determined from
the Boltzmann equation \cite{GyS95,vNyE01,ByP04}
\begin{equation}
\left( \partial _{t}+{\bm v} \cdot {\bm \nabla}\right) f=J[f,f],
\label{2.1}
\end{equation}
where $J$ is the inelastic Boltzmann collision operator defined by
\begin{equation}
J[X,Y]\equiv \sigma ^{d-1}\int d{\bm v}_{1}\int d\widehat{\bm
\sigma}\, \Theta (\widehat{\bm \sigma}\cdot {\bm g}) \widehat{\bm
\sigma}\cdot {\bm g} \left[ \alpha ^{-2}X({\bm r},{\bm v}^{\prime
},t,)Y({\bm r},{\bm v}_{1}^{\prime },t)-X({\bm r},{\bm v},t)Y({\bm
r},{\bm v}_{1},t)\right] , \label{2.2}
\end{equation}
for arbitrary functions $X({\bm v})$ and $Y({\bm v})$. Here,
$\widehat{\bm \sigma}$ is a unit vector along the line joining the
centers of the colliding pair, $\Theta $ is the Heaviside step
function, and ${\bm g}={\bm v}-{\bm v}_{1}$. The primes on the
velocities denote the initial values $\left\{ {\bm v}_{1}^{\prime
},{\bm v} _{2}^{\prime }\right\} $ that lead to $\left\{ {\bm
v}_{1},{\bm v} _{2}\right\} $ following a ``restituting'' binary
collision,
\begin{equation}
{\bm v}^{\prime }={\bm v}-\frac{1}{2}(1+\alpha ^{-1})(\widehat{
\bm \sigma }\cdot {\bm g})\widehat{\bm \sigma}, \quad \quad {\bm
v}_{1}^{\prime }={\bm v}_{1}+\frac{1}{2}(1+\alpha ^{-1})(
\widehat{\bm \sigma}\cdot {\bm g})\widehat{\bm \sigma}.
\label{2.3}
\end{equation}
The usual Boltmann collision operator is recovered from Eq.\
(\ref{2.2}) in the elastic limit $\alpha =1$.

The macroscopic variables of interest are the hydrodynamic fields,
i.e.  the density $n({\bm r},t)$, the flow velocity ${\bm u}({\bm
r},t)$, and the (granular) temperature $T({\bm r},t)$. They are
defined as moments of the solution to the Boltzmann equation,
\begin{equation}
\left(
\begin{array}{c}
n({\bm r},t) \\
n({\bm r},t){\bm u}({\bm r},t) \\
\frac{d}{2}n({\bm r},t)T({\bm r},t)
\end{array}
\right) =\int d{\bm v}\left(
\begin{array}{c}
1 \\
{\bm v} \\
\frac{1}{2}m\left( {\bm v}-{\bm u}\right)^{2}
\end{array}
\right) f({\bm r},{\bm v},t).  \label{2.4}
\end{equation}
An exact set of equations for these variables is obtained from the
following properties of the collision operator
\begin{equation}
\int d{\bm v}\left(
\begin{array}{c}
1 \\
{\bm v} \\
\frac{1}{2}m\left( {\bm v}-{\bm u}\right) ^{2}
\end{array}
\right) J[f,f]=\left(
\begin{array}{c}
0 \\
\mathbf{0} \\
-\frac{d}{2}nT\zeta
\end{array}
\right).  \label{2.5}
\end{equation}
The first two equations follow from conservation of mass and
momentum in the particle collisions. The last equation reflects
the loss of energy in collisions due to the inelasticity. This
appears through the ``cooling rate'' $\zeta ({\bm r},t) \geq 0$,
defined by this equation.

The macroscopic balance equations resulting from the above
properties are:
\begin{equation}
\partial _{t}n+{\bm \nabla}\cdot \left( n {\bm u}\right) =0,
\label{2.6}
\end{equation}
\begin{equation}
\left( \partial _{t}+ {\bm u} \cdot {\bm \nabla}\right) {\bm
u}+(mn)^{-1}\left[{\bm \nabla} (nT) + {\bm \nabla} \cdot
\mathsf{\Pi}\right] =0, \label{2.7}
\end{equation}
\begin{equation}
\left( \partial _{t}+ {\bm u} \cdot {\bm \nabla} + \zeta \right) T
+ \frac{2}{nd}\left( nT {\bm \nabla} \cdot {\bm u}+\mathsf{\Pi}:
{\bm \nabla} {\bf u}+{\bm \nabla}\cdot {\bm q}\right) =0.
\label{2.8}
\end{equation}
The functionals giving the dissipative part of the pressure
tensor, $\mathsf{\Pi}$, and the heat flux, ${\bf q}$, are also
moments of the solution to the Boltzmann equation,
\begin{equation}
\mathsf{\Pi}(\mathbf{r},t)=\int d{\bm v}\, m \left( {\bm V}{\bm
V}-\frac{V^{2}}{d}\mathsf{I}\right) \,f({\bm r},{\bm v},t),
\label{2.9}
\end{equation}
\begin{equation}
{\bm q}({\bm r},t)=\int d{\bm v}\, \left(
\frac{m}{2}V^{2}-\frac{d+2}{2}T\right) {\bm V}\,f({\bm r},{\bm
v},t), \label{2.10}
\end{equation}
where $\mathsf{I}$ is the unit tensor of dimension $d$ and ${\bm
V} \equiv {\bm v}-{\bm u}$ the so-called peculiar velocity.
Equations (\ref {2.6})-(\ref{2.8}) are the basis for a
hydrodynamic description, once $ \mathsf{\Pi}({\bm r},t)$, ${\bm
q}({\bm r},t)$, and $\zeta ({\bm r},t)$ are specified. These can
be obtained from their definitions using a solution of the
Boltzmann equation generated by the Chapman-Enskog method. This
method assumes the existence of a solution whose space and time
dependence is given entirely through the hydrodynamic fields and
their gradients (i.e., a ``normal'' solution). As a result,
$\mathsf{\Pi}({\bm r},t)$ and ${\bm q}( {\bm r},t)$ are given in
terms of these variables, and Eqs. (\ref{2.6})-(\ref {2.8}) become
a closed set of hydrodynamic equations. The primary results of
this method are recalled in Appendix \ref{ap1}. To leading order
in the spatial gradients, the dissipative fluxes are found to be
given by \cite{BDKyS98,ByC01}
\begin{equation}
\mathsf{\Pi}_{ij}({\bm r},t)= -\eta \left( \nabla_{i}
u_{j}+\nabla_{j} u_{i}-\frac{2}{d}\delta _{ij} {\bm \nabla}\cdot
{\bm u}\right),   \label{2.11}
\end{equation}
\begin{equation}
\mathbf{q}(\mathbf{r},t) = -\kappa {\bm \nabla}T-\mu {\bm
\nabla}n. \label{2.12}
\end{equation}
For the cooling rate the result is:
\begin{equation}
\zeta ({\bm r},t)= n\sigma ^{d-1}\left( \frac{2T}{m}
\right)^{1/2}\zeta ^{\ast }_{0}+\zeta _{1}\nabla ^{2}T+\zeta
_{2}\nabla ^{2}n+  \text{bilinear in } {\bm \nabla} n,{\bm \nabla}
T,{\bm \nabla} {\bm u} \text{ terms}, \label{2.13}
\end{equation}
with $\zeta ^{\ast}_{0}$ being a dimensionless positive constant
proportional to $ \left( 1-\alpha ^{2}\right)$. The explicit forms
for the shear viscosity $ \eta $, the thermal conductivity $\kappa
$, and the new transport coefficients $\mu$, $\zeta _{1}$, and
$\zeta _{2}$, peculiar to granular gases, are given in Appendix
\ref{ap1}. The nonlinear contributions to $\zeta (t)$ indicated in
(\ref{2.13}) play no role in the following linear analysis and
will be not discussed further in this paper.

Equations (\ref{2.6})-(\ref{2.8}) together with the ``constitutive
relations'' (\ref {2.11})-(\ref{2.13}) are the Navier-Stokes
hydrodynamic equations for a granular gas. A special solution of
these equations is that for a homogeneous state, characterized by
a constant density $n_{hcs}$, a vanishing velocity flow
$\mathbf{u}_{hcs}=0$, and a uniform temperature $T_{hcs}(t)$
determined from
\begin{equation}
\partial _{t}T_{hcs}\left( t\right) =-n_{hcs} \sigma ^{d-1}\left( \frac{2}{m}
\right)^{1/2}\zeta^{\ast}_{0}T_{hcs}^{3/2}\left( t\right) .
\label{2.14}
\end{equation}
This is referred to as the homogeneous cooling state (HCS), since
the only macroscopic dynamics is the monotonic decrease in the
temperature field. It is easily seen that the HCS as defined above
is also a solution of the exact balance equations, taking into
account that $\zeta_{hcs} \propto n_{hcs} \sigma^{d-1} (
T_{hcs}/m)^{1/2}$ on dimensional grounds. In the following, the
solution to the Boltzmann equation will be considered for small
perturbations of the HCS, and it will be useful to have the
corresponding results from Navier-Stokes hydrodynamics. The
linearized Navier-Stokes equations have time dependent
coefficients, since the distribution function of the reference HCS
depends on time. This complication can be overcome by the
introduction of suitable dimensionless space and time scales, as
well as scaled hydrodynamic fields. Thus we define ${\bm r}^{*}$
and $s$ by
\begin{equation}
{\bm r}^{*}=\frac{\bm r}{\ell}, \quad ds=\frac{v_{hcs}(t)}{\ell}\,
dt. \label{2.15}
\end{equation}
Here, $v_{hcs}\left( t\right) \equiv \left[ 2 T_{hcs}(t)/m
\right]^{1/2} $ is the thermal velocity in the HCS, and $\ell
\equiv (n_{hcs}\sigma^{d-1})^{-1}$ is proportional to the mean
free path of the particles. The dimensionless fields $\delta
y_{j}({\bm r}^{*},s)$ are chosen as
\begin{equation}
\rho ({\bm r}^{*},s)=\frac{n({\bm r},t)-n_{hcs}}{n_{hcs}}, \quad
\theta ({\bm r}^{*},s)=\frac{T({\bm r},t)-T_{hcs}(t)}{T_{hcs}(t)},
\quad {\bm \omega}({\bm r^{*}},s)= \frac{{\bm u}({\bm
r},t)}{v_{hcs}(t)}. \label{2.16}
\end{equation}
Besides, since the equations are linear and the reference state is
homogeneous, it is sufficient to consider a single Fourier mode,
\begin{equation}
\delta y_{j }\left({\bm r}^{\ast },s\right) =e^{i {\bm k} \cdot
{\bm r} ^{\ast }}\delta \widetilde{y}_{j }\left( {\bm k},s\right).
\label{2.17}
\end{equation}
The velocity components are chosen as a longitudinal component
relative to ${\bm k}$, $\widetilde{\omega}_{\parallel}\left( {\bm
k},s\right) =\widehat{\bm k} \cdot \widetilde{\bm \omega}\left(
{\bm k},s \right) $, and $d-1$ transverse components
$\widetilde{\omega}_{\perp,i}({\bm k},s)=\widehat{\bm e}^{i} \cdot
\widetilde{\bm \omega}({\bm k},s)$, where $\{ \widehat{\bm
k}\equiv {\bm k}/k,\widehat{\bm e}^{(i)}, i=1, \ldots, d-1 \}$ are
a set of $d$ pairwise orthogonal unit vectors . The dimensionless
linear Navier-Stokes equations then become a system of ordinary
differential equations with constant coefficients that can be
expressed in the compact form \cite{BDKyS98}
\begin{equation}
\partial _{s} \delta \widetilde{\bm y}({\bm k},s)+\mathsf{K}\left( k\right)
\cdot  \delta \widetilde{\bm y}({\bm k},s)=0, \label{2.18}
\end{equation}
where we use a $d+2$ dimensional space representation with $\delta
\widetilde{\bm y}({\bm k},s)$ being the vector
\begin{equation}
\delta \widetilde{\bm y}({\bm k},s) = \left(
\begin{array}{l}
\widetilde{\rho} ({\bm k},s) \\
\widetilde{\theta} ({\bm k},s) \\
\widetilde{\omega}_{\parallel}({\bm k},s) \\
\widetilde{\bm \omega}_{\perp}({\bm k},s)
\end{array}
\right), \label{2.18a}
\end{equation}
and $\widetilde{\bm \omega}_{\perp}({\bm k},s)$ denoting the
vector formed by the $d-1$ components
$\widetilde{\omega}_{\perp,i} ({\bm k},s)$. The matrix
$\mathsf{K}\left( k\right) $ is block diagonal, expressing the
decoupling of transverse and longitudinal modes,
\begin{equation}
\mathsf{K}(k) = \left(
\begin{array}{cc}
\mathsf{K}_{1} & 0 \\
0 & \mathsf{K}_{2}
\end{array}
\right), \label{2.18b}
\end{equation}
\begin{equation}
\mathsf{K}_{1}(k) = \left(
\begin{array}{ccc}
0 & 0 & ik \\
\zeta^{\ast}_{0}+\left( \mu ^{\ast }-\zeta _{2}^{\ast }\right)
k^{2} & \frac{\zeta_{0}^{\ast }}{2}+\left( \kappa ^{\ast }-\zeta
_{1}^{\ast }\right) k^{2} &
\frac{2i}{d}k \\
\frac{i}{2}k & \frac{i}{2}k &
-\frac{\zeta^{*}_{0}}{2}+\frac{2(d-1)}{d}\eta ^{\ast }k^{2}
\end{array}
\right),  \label{2.19}
\end{equation}
\begin{equation}
\mathsf{K}_{2}(k)= - \left( \frac{\zeta^{*}_{0}}{2} - \eta^{*}
k^{2} \right) \mathsf{I}. \label{2.19a}
\end{equation}
In the above expressions, $\mathsf{K}_{2}$ and $\mathsf{I}$ are
matrices of dimension $d-1$. The dimensionless transport
coefficients used in the above equations are defined by
\begin{equation}
\eta ^{\ast }=\frac{\eta}{\ell m n_{hcs} v_{hcs}}, \quad
\kappa^{\ast}=\frac{2 \kappa}{d \ell n_{hcs} v_{hcs}}, \quad
\mu^{*} = \frac{ 2 \mu}{ d \ell T_{hcs} v_{hcs}}, \label{2.19b}
\end{equation}
\begin{equation}
\zeta _{1}^{\ast }=\frac{T_{hcs} \zeta_{1}}{ \ell v_{hcs} }, \quad
\zeta _{2}^{\ast }=\frac{n_{hcs}\zeta_{2}}{ \ell v_{hcs} }\,.
\label{2.19c}
\end{equation}

The formal solution to the initial value problem (\ref{2.18}) is
\begin{equation}
\delta \widetilde{\bm y}({\bm k},s)= e^{-\mathsf{K}\left( k\right)
s} \cdot \delta \widetilde{\bm y}({\bm k},0). \label{2.20}
\end{equation}
The eigenvalues and eigenvectors of the generator $\mathsf{K}$ for
this dynamics define the $d+2$ Navier-Stokes order hydrodynamic
modes. They are given by the solutions of the equation
\begin{equation}
\mathsf{K}(k) \cdot {\bm \varphi}_{j}({\bm k})=\lambda _{j}(k){\bm
\varphi}_{j }({\bm k}), \quad j=1, \ldots, d+2. \label{2.21}
\end{equation}
A simple calculation provides the expressions for the eigenvalues
to order $k^{2}$. They are given by
\begin{equation*}
\lambda_{1}(k) =\frac{k^{2}}{\zeta^{\ast}_{0}}, \quad \lambda
_{2}(k)=\frac{\zeta^{*}_{0}}{2}- \left( \frac{d+1}{d
\zeta_{0}^{*}} -\kappa ^{\ast }+\zeta _{1}^{\ast } \right) k^{2},
\end{equation*}
\begin{equation}
\lambda_{\parallel}(k) =-\frac{\zeta^{*}_{0}}{2}+\left[ \frac{1}{d
\zeta_{0}^{*}}+ \frac{2(d-1) \eta^{*}}{d}\right] k^{2}, \quad
\lambda _{\perp}(k)=-\frac{\zeta^{*}_{0}}{2}+ \eta ^{\ast }k^{2},
\label{2.23}
\end{equation}
the eigenvalue $\lambda_{\perp}(k)$ being $(d-1)$-fold degenerate.
The corresponding eigenvectors to leading order in $k$ are
\begin{equation*}
{\bm \varphi} _{1}({\bm k})=\left(
\begin{array}{c}
1 \\
-2 \\
0 \\
{\bm 0}
\end{array}
\right) ,\quad {\bm \varphi}_{2}({\bm k})=\left(
\begin{array}{c}
0 \\
1 \\
0 \\
{\bm 0}
\end{array}
\right) ,\quad {\bm \varphi} _{\parallel}({\bm k})=\left(
\begin{array}{c}
0 \\
0 \\
1 \\
{\bm 0}
\end{array}
\right) ,
\end{equation*}
\begin{equation}
{\bm \varphi}_{\perp,i}({\bm k})=\left(
\begin{array}{c}
0 \\
0 \\
0 \\
{\widehat{\bm i}}
\end{array}
\right).   \label{2.24}
\end{equation}
Here,
\begin{itemize}
\item ${\bm 0}=0$, and $\widehat{\bm i}=\widehat{\bm 1}=1$, for
$d=2$.

\item ${\bm 0}= \left( \begin{array}{c} 0 \\ 0 \end{array}
\right)$, \quad $\widehat{\bm 1}= \left( \begin{array}{c} 1 \\ 0
\end{array} \right)$, \quad and $\widehat{\bm 2}= \left( \begin{array}{c}
0 \\ 1 \end{array} \right)$, for $d=3$.

\end{itemize}

The first of these modes is excited by the condition
$\widetilde{\theta} \left({\bm k},0
\right)=-2\widetilde{\rho}({\bm k},0)$ at zero flow velocity. The
second is produced by a temperature perturbation at constant
density and also zero velocity, while the third one corresponds to
a longitudinal velocity perturbation at constant temperature and
density. There is a $(d-1)$-fold degeneracy for the shear modes of
eigenvalue $\lambda_{\perp}$. These diffusive modes are excited by
a perturbation of the velocity field in the transverse plane
orthogonal to ${\bm k}$.

It should be noted that while the above analysis is restricted to
the Navier-Stokes equations, derived from the Boltzmann equation
by the Chapman-Enskog method, the eigenvalues and eigenvectors to
order $k$ follow more generally from the exact macroscopic balance
equations and do not depend on the approximate constitutive
equations (\ref{2.11}) and (\ref{2.12}). The hydrodynamic modes
sought by kinetic theory in the subsequent sections can therefore
be defined as those excitations due to small perturbations which
agree with the above in the long wavelength limit. Analyticity
then allows extension of that identification to shorter
wavelengths. A consistency check of the Chapman-Enskog method is
agreement with the above results at order $k^{2}$. However, the
concept of hydrodynamic modes in this context does not require the
validity of the Chapman-Enskog method nor the limitation to the
Navier-Stokes approximation.

\section{Linear Boltzmann Equation}
\label{s3}

A more complete and accurate description of the response to small
spatial perturbations of the density, temperature, and flow
velocity is obtained directly from the Boltzmann equation.
Consider first an isolated system. As already indicated in the
previous section, the exact balance equations have a solution
describing the HCS, with a monotonically decreasing temperature
obeying Eq.\ (\ref{2.14}). The solution to the Boltzmann equation
corresponding to this macroscopic state is characterized by the
scaling form \cite{GyS95}
\begin{equation}
f_{hcs}({\bm v},t)=n_{hcs}v_{hcs}^{-d}(t)\phi \left( V^{\ast
}\right), \label{3.2}
\end{equation}
with
\begin{equation} {\bm V}^{\ast }=\frac{\bm V}{v_{hcs}(t)}=
\frac{{\bm v}-{\bm u}_{hcs}}{v_{hcs}(t)}. \label{3.1}
\end{equation}
Substitution into the Boltzmann equation gives
\begin{equation}
\frac{\zeta _{hcs}}{2} \frac{\partial}{\partial {\bm V}}\cdot
\left( \mathbf{V} f_{hcs}\right) =J[f_{hcs},f_{hcs}]. \label{3.3}
\end{equation}
For later convenience, a constant velocity $ \mathbf{u}_{hcs}$ for
the system as a whole has been included, although this can always
be removed by means of a Gallilean transformation. Then, in the
following it will be considered that ${\bm u}_{hcs}=0$, unless it
be explicitly established otherwise. An exact and explicit
solution of Eq.\ (\ref{3.3}) is not known yet, but the behavior of
$\phi $ at large and small velocities has been determined
\cite{vNyE01} and the results obtained by the direct simulation
Monte Carlo method strongly supports the existence of such a
scaling form \cite{BRyC96}.

The HCS distribution function is a ``universal'' homogeneous
solution in the same sense as the Maxwellian for elastic
collisions. An arbitrary homogeneous state is expected to approach
the HCS after a few collisions. Therefore, in discussing response
of any homogeneous state to small spatial perturbations, it is
sufficient to consider the HCS as the reference state. All the
other cases will simply induce additional short time transients.

Consider then small spatial perturbations of the HCS
\begin{equation}
f({\bm r},{\bm v},t)=f_{hcs}({\bm v},t)\left[ 1+\Delta ({\bm
r},{\bm v, }t)\right], \quad |\Delta({\bm r},{\bm v},t)| \ll 1.
\label{3.4}
\end{equation}
To linear order in $\Delta ({\bm r},{\bm v,}t)$, the Boltzmann
equation becomes
\begin{equation}
\left[ \partial _{t}+{\bm v} \cdot {\bm \nabla} + L(t)\right]
(f_{hcs} \Delta) =0,  \label{3.5}
\end{equation}
where $L(t)$ is the linearized Boltzmann collision operator given
by
\begin{equation}
L(t)X({\bm v})=-J[f_{hcs},X]-J[X,f_{hcs}], \label{3.6}
\end{equation}
for arbitrary $X({\bm v})$. Just as for the analysis of the
Navier-Stokes equations in the previous section, the above linear
kinetic equation takes a simpler form when expressed in terms of
the dimensionless variables defined in Eq.\ (\ref{2.15}), and
considering a single Fourier mode
\begin{equation}
\Delta ({\bm r},{\bm v},t)=e^{i{\bm k} \cdot {\bm r}^{\ast
}}\widetilde{\Delta} ({\bm k},{\bm v}^{\ast},s),  \label{3.7}
\end{equation}
where ${\bm v}^{\ast }= {\bm v}/  v_{hcs}(t)$. Then, Eq.\
(\ref{3.5}) becomes
\begin{equation}
\left( \partial _{s}+i {\bm k} \cdot {\bm v}^{\ast
}+\mathcal{L}^{\ast }\right) \widetilde{\Delta}({\bm k},{\bm
v}^{\ast },s)=0. \label{3.8}
\end{equation}
The dimensionless operator $\mathcal{L}^{*}$ is now time
independent and it is given by
\begin{equation}
\mathcal{L}^{\ast }X \equiv \frac{\zeta^{*}_{0}}{2} \phi ^{-1}
\frac{\partial}{\partial {\bm v}^{*}} \cdot \left({\bm v}^{\ast
}\phi X\right)  +L^{\ast }X, \label{3.9}
\end{equation}
where $L^{*}$ is the dimensionless linear Boltzmann collision
operator
\begin{equation}
L^{\ast }X \equiv -\phi ^{-1}\left( J^{*}[\phi ,\phi X]+J^{*}
[\phi X,\phi ]\right), \label{3.10}
\end{equation}
\begin{equation}
J^{*}[X,Y] \equiv \int d{\bm v}^{*}_{1} \int d \widehat{\bm
\sigma}\, \Theta (\widehat{\bm \sigma} \cdot {\bm g}^{*}  )
\widehat{\bm \sigma} \cdot {\bm g}^{*} \left[ \alpha^{-2} X({\bm
v}^{*\prime}) Y({\bm v}^{* \prime}_{1})-X({\bm v}^{*})Y({\bm
v}^{*}_{1}) \right]. \label{3.10a}
\end{equation}
Here, ${\bm v}^{* \prime}$ and ${\bm v}^{* \prime}_{1}$ are
related with ${\bm v}^{*}$ and ${\bm v}^{*}_{1}$ by Eqs.\
(\ref{2.3}). The operator $\mathcal{L}^{*}$ differs from the
linearized Boltzmann collision operator $L^{*}$ by terms
representing the cooling effects of the inelastic collisions. The
latter arise because the derivative with respect to $s$ is taken
at constant ${\bm v}^{\ast }$ rather than ${\bm v}$.

Solutions to the dimensionless, linear kinetic equation
(\ref{3.8}) for $ \widetilde{\Delta}$ are sought in a Hilbert
space whose scalar product is defined by
\begin{equation}
\left( X,Y\right) =\int d\mathbf{v}^{\ast }\, \phi \left( v^{\ast
}\right) X^{\dagger }\left( {\bm v}^{\ast }\right) Y({\bm v}^{\ast
}), \label{3.11}
\end{equation}
with the dagger denoting complex conjugation. The formal solution
to the kinetic equation is
\begin{equation}
\widetilde{\Delta} \left( {\bm k},{\bm v}^{\ast },s\right) =
\frac{1}{2 \pi i} \oint dz\, e^{-zs}\mathcal{R}(z)\Delta ^{\ast
}\left( {\bm k}, {\bm v}^{\ast },0\right) , \label{3.12}
\end{equation}
\begin{equation}
\mathcal{R}(z) \equiv \left( z-i {\bm k} \cdot {\bm v}^{\ast
}-\mathcal{L}^{\ast }\right) ^{-1},  \label{3.13}
\end{equation}
where the contour encloses the entire spectrum of $i {\bm k} \cdot
{\bm v}^{\ast }+\mathcal{L}^{\ast }$ , both point and residual,
counterclockwise. It is important to realize that \emph{ all the
linear excitations of the granular gas are determined from this
spectrum. } This formulation of the problem for small spatial
perturbations provides a precise context for addressing many
questions regarding hydrodynamics for a granular gas. The
existence of the hydrodynamic modes and their role relative to
other dynamical processes are determined by the characterization
of the above spectrum. To see how this occurs, it is useful to
recall briefly the status of the corresponding problem for gases
with elastic collisions \cite{Mc89,Scharf}. In that case, it has
been proved that the hydrodynamic modes exist as $d+2$ poles
located at the origin in the long wavelength limit and
corresponding to the local conserved quantities. Furthermore, the
remainder of the spectrum is bounded away from these poles, and
the spectrum is analytic in ${\bm k}$ \ about ${\bm k}=0$, so this
isolation of the hydrodynamic modes is preserved at finite
wavelengths.

\section{Existence of Hydrodynamic Modes}
\label{s4}

The spectrum of $i{\bm k} \cdot {\bm v}^{\ast }+\mathcal{L}^{\ast
}$ is expected to be quite complex, based on the special case of
elastic collisions, with points spectra, continua, and limit
points. The hydrodynamic excitations, whenever they exist, are
part of the point spectrum so in order to investigate them it
suffices to consider the eigenvalue problem
\begin{equation}
\left( i {\bm k} \cdot {\bm v}^{\ast }+\mathcal{L}^{\ast }\right)
\Psi _{i}=\lambda _{i}\left( k\right) \Psi _{i}.  \label{4.1}
\end{equation}
The search for hydrodynamic excitations can be carried out by
assuming they are analytic in $k$ and looking first for the ${\bm
k}=0$ solutions of Eq.\ (\ref{4.1}). The practical issue of
constructing these modes at finite $k$ is addressed in the next
section.

The central idea for constructing the hydrodynamic eigenvalues and
eigenvectors at $k=0$ is the note that the HCS is parameterized by
the hydrodynamic fields $n_{hcs},$ $T_{hcs}$, and ${\bm u}_{hcs}$,
which is now considered different from zero, as discussed at the
beginning of the previous section. Therefore, differentiating the
Boltzmann equation for the distribution function of the HCS, Eq.\
(\ref{3.3}), with respect to these fields gives exact properties
of the linearized Boltzmann collision operator. For example,
\begin{equation}
\frac{\partial }{\partial n_{hcs}} \left\{ \frac{\zeta_{hcs}}{2}\,
\frac{\partial}{\partial \bm V}\ \cdot \left(
\mathbf{V}f_{hcs}\right) -J[f_{hcs},f_{hcs}]\right\} =0
\end{equation}
gives directly
\begin{equation}
\frac{1}{2}\frac{\partial \zeta _{hcs}}{\partial n_{hcs}}
\frac{\partial}{\partial {\bm V}} \cdot \left(
\mathbf{V}f_{hcs}\right) +\frac{\zeta_{hcs}}{2}\,
\frac{\partial}{\partial {\bm V}}\, \cdot \left( {\bm V}
\frac{\partial f_{hcs}}{
\partial n_{hcs}}\right) + L\frac{\partial f_{hcs}}{\partial
n_{hcs}}=0.  \label{4.2}
\end{equation}
The derivatives of $\zeta _{hcs}$  and $f_{hcs}$ are easily
calculated from the properties $ \zeta _{hcs}\propto n_{hcs}$ and
$f_{hcs} \propto n_{hcs}$. In terms of the dimensionless
variables, and setting $\mathbf{u}_{hcs}=0$, Eq.\ (\ref{4.2})
becomes
\begin{equation}
\mathcal{L}^{\ast }1=-\frac{\zeta_{0}^{*}}{2}\, \phi ^{-1}
\frac{\partial}{\partial {\bm v}^{*}}\, \cdot \left( {\bm v}^{\ast
}\phi \right) . \label{4.3}
\end{equation}

Next, calculate the derivative of Eq.\ (\ref{3.3}) with respect to
$T_{hcs}$,
\begin{equation}
\frac{1}{2}\frac{\partial \zeta _{hcs}}{\partial T_{hcs}}\,
\frac{\partial}{\partial {\bm V}}\, \cdot \left( {\bm V}
f_{hcs}\right) +\frac{\zeta_{hcs}}{2} \frac{\partial}{\partial
{\bm V}}\, \cdot \left( \mathbf{V}\frac{\partial f_{hcs}}{
\partial T_{hcs}}\right) +L\frac{\partial f_{hcs}}{\partial
T_{hcs}}=0.  \label{4.4}
\end{equation}
Since $f_{hcs}$ has the scaling form (\ref{3.2}), it is
\begin{equation}
\frac{\partial f_{hcs}}{\partial T_{hcs}}=-
\frac{f_{hcs}}{2T_{hcs}}\left( d+ {\bm V} \cdot \frac{\partial \ln
f_{hcs}}{\partial {\bm V}} \right), \label{4.4a}
\end{equation}
and, taking into account that $\zeta_{hcs} \propto T_{hcs}^{1/2}$,
Eq. (\ref{4.4}) becomes
\begin{eqnarray}
\frac{\zeta_{hcs}}{2} \frac{\partial}{\partial {\bm V}} \cdot
\left( {\bm V} f_{hcs}\right)  & - & \frac{\zeta_{hcs}}{2}
\frac{\partial}{\partial {\bm V}} \cdot \left[ {\bm
V}f_{hcs}\left( d+{\bm V} \cdot \frac{\partial \ln
f_{hcs}}{\partial {\bm V}}\right) \right] \nonumber \\
& - & L\left[ f_{hcs} \left( d+{\bm V} \cdot \frac{\partial \ln
f_{hcs}}{\partial {\bm V}} \right) \right] =0. \label{4.5}
\end{eqnarray}
Setting ${\bm u}_{hcs}=0$ and transforming to dimensionless
variables, the above equation yields
\begin{equation}
\mathcal{L}^{\ast }\left( d+{\bm v}^{\ast } \cdot \frac{\partial
\ln \phi}{\partial {\bm v^{*}}} \right) =\frac{\zeta^{*}_{0}}{2}
\left( d+ {\bm v}^{*} \cdot \frac{\partial \ln \phi}{\partial {\bm
v}^{*}} \right) . \label{4.6}
\end{equation}
Finally, differentiating Eq.\ (\ref{3.3}) with respect to
$\mathbf{u}_{hcs}$, afterwards setting $\mathbf{u}_{hcs}=0$, and
introducing dimensionless variables leads in a similar way to
\begin{equation}
\mathcal{L}^{\ast }\left( \frac{\partial \ln \phi}{\partial {\bm
v}^{*}} \right)=  - \frac{\zeta_{0}^{*}}{2} \left( \frac{\partial
\ln \phi}{\partial {\bm v}^{*}} \right). \label{4.7}
\end{equation}

The $d+2$ equations (\ref{4.3}), (\ref{4.6}), and (\ref{4.7})
provide exact properties of $\mathcal{L}^{\ast }$. In fact, Eqs.\
(\ref{4.6}) and (\ref{4.7}) are of the form of the eigenvalue
problem to be solved at $k=0$, with eigenvalues given by $ \zeta
_{0}^{*}/2$ and $- \zeta_{0}^{*}/2$, respectively. It is
straightforward to construct linear combinations of (\ref{4.3})
and (\ref{4.6}) to obtain an additional eigenvalue and
eigenvector. The results can then be expressed as \cite{DyB03}
\begin{equation}
\mathcal{L}^{\ast }\Psi _{i}(0)=\lambda _{i}\left( 0\right) \Psi
_{i}(0), \quad i=1, \ldots, d+2,  \label{4.8}
\end{equation}
with
\begin{equation}
\left\{ \lambda _{i}(0)\right\} =\left\{
0,\frac{\zeta^{*}_{0}}{2},-\frac{\zeta^{*}_{0}}{2} \right\},
\label{4.9}
\end{equation}
\begin{equation}
\left\{ \Psi _{i}(0)\right\}  =\left\{ d+1+ {\bm v}^{*} \cdot
\frac{\partial \ln \phi}{\partial {\bm v}^{*}}, -d- {\bm v}^{*}
\cdot \frac{\partial \ln \phi}{\partial {\bm v}^{*}},
-\widehat{\bm k} \cdot \frac{\partial \ln \phi}{\partial {\bm
v}^{*}}, -\frac{\partial \ln \phi}{\partial {\bm v}^{*}_{\perp}}
\right\}. \label{4.10}
\end{equation}
The eigenvalue $-\zeta_{0}^{*}/2$ is $d$-fold degenerated. For
convenience for the perturbation calculation to be carried out in
the following section, the subspace associated to it has been
rearranged. The velocity ${\bm v}^{*}$ has been decomposed into
its component in the direction of ${\bm k}$, $v^{*}_{\parallel}
=\widehat{\bm k} \cdot {\bm v}^{*}$, and the remaining $d-1$ ones,
forming with it a set of $d$ pairwise orthogonal components, i.e.
${\bm v}^{*}_{\perp}$ is defined by the $d-1$ components ${\bm
v}^{*} \cdot \widehat{\bm e}^{(i)}$. When appropriate, the set of
the $d$ eigenfunctions associated with the eigenvalue
$-\zeta_{0}^{*}/2$ will be denoted by ${\bm \Psi}_{3}(0)$.

Clearly, the above are the long wavelength limit of the
hydrodynamic modes defined by Eq.\ (\ref{2.23}). This is a the
first primary result of the analysis developed here, demonstration
of the existence of hydrodynamic excitations in the spectrum of
the linearized Boltzmann equation. The results are exact and apply
for arbitrary degree of dissipation.

\begin{figure}
\includegraphics[scale=0.5,angle=-90]{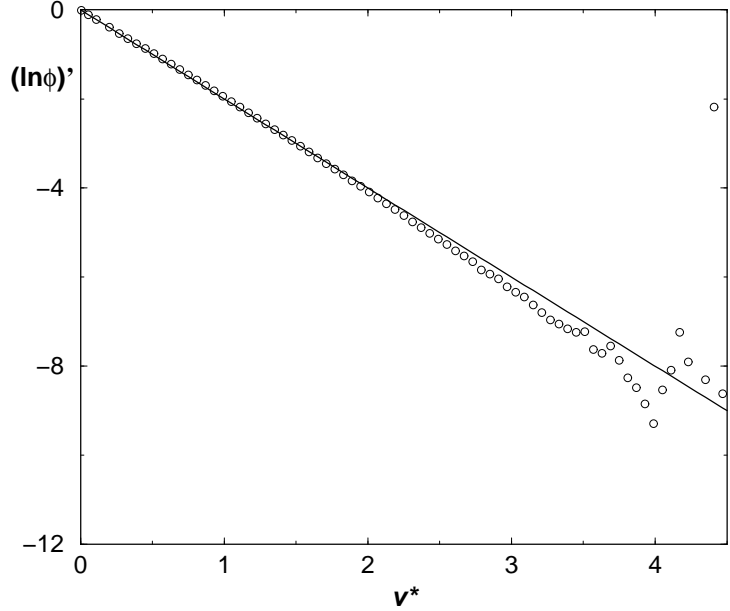}
\caption{Plot of $(\ln \phi)^{\prime} \equiv \partial \ln
\phi/\partial v^{*}$ as a function of $v^{*}$ for $d=3$ and
$\alpha=0.95$. The circles are the numerical derivative of the
DSMC results and the solid line is the Gaussian. Quantities are
measured in the dimensionless units defined in the text.
\label{fig1} }
\end{figure}

\begin{figure}
\includegraphics[scale=0.5,angle=-90]{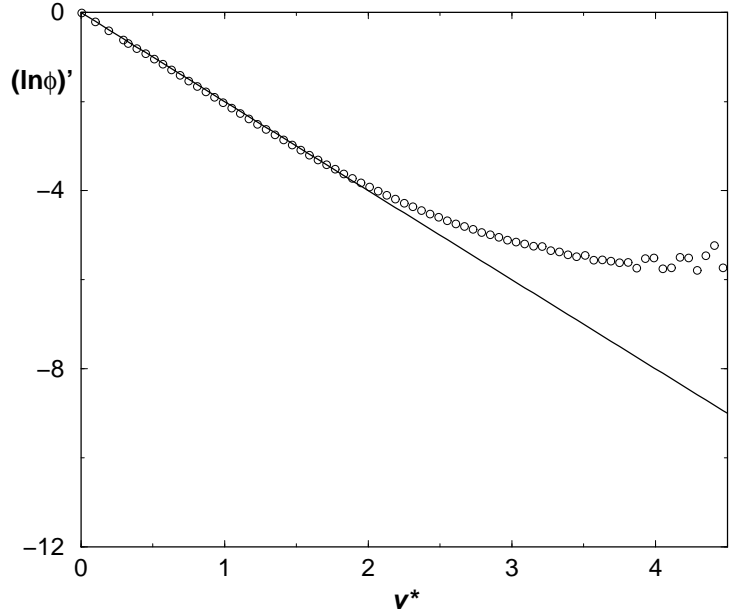}
\caption{The same as Fig.\ \ref{fig1} but for $\alpha=0.6$.
\label{fig2} }
\end{figure}

These long wavelength eigenfunctions are determined from the HCS
distribution which depends only on the magnitude of $v^{*}$, so
the terms on the right side of Eq.\ (\ref{4.10}) are all
determined from $\partial \ln \phi \left( v^{*} \right) / \partial
v^{*}$. In the elastic limit, $\phi \left( v^{*} \right) $ becomes
Gaussian, and the hydrodynamic eigenfunctions become linear
combinations of $ 1,$ ${\bm v}^{*}$, and $v^{* 2}$, i.e. of the
summational invariants for the conservation laws of mass,
momentum, and energy, as expected. For inelastic collisions, the
eigenfunctions are quite different, particularly at large $v^{*}$
due to an overpopulation in the HCS distribution relative to the
Gaussian. This is illustrated in Figs.  \ref{fig1} and \ref{fig2},
where $\partial \ln \phi \left( v\right)/\partial v^{*} $ has been
obtained from Direct Simulation Monte Carlo solution of the
Boltzmann equation for the HCS \cite{Bi94}.

The set of eigenfunctions $\Psi_{i}(0)$ span a $d+2$ dimensional
subspace of the Hilbert space, but they are not orthogonal.
Consistently, it is easily verified that the operator
$\mathcal{L}^{\ast }$ is not Hermitian. Therefore, it is useful to
introduce a biorthogonal set of functions $\Phi_{i}, i=1, \ldots,
d+2$, with the requirement
\begin{equation}
\left( \Phi _{i},\Psi _{j}(0)\right) =\delta _{ij}.  \label{4.11}
\end{equation}
To identify the appropriate set of functions, first note that
\begin{equation}
\left( 1,\mathcal{L}^{\ast }X\right) =0, \quad \left( {\bm v}
^{\ast },\mathcal{L}^{\ast }X\right) =-\frac{\zeta^{*}_{0}}{2}
\left( {\bm v}^{*},X \right), \label{4.12}
\end{equation}
as a consequence of the number of particles and momentum
conservation in the moment equations (\ref{2.5}). Therefore, $1$
and ${\bm v}^{*}$ are eigenfunctions of the adjoint of
$\mathcal{L}^{\ast }$,  $\mathcal{L}^{* \dagger }$, with
eigenvalues $0$ and -$\zeta^{\ast}_{0}/2$, respectively. This
gives the set
\begin{equation}
\left\{ \Phi _{i}\right\} =\left\{ 1,\Phi _{2},\widehat{\bm k}
\cdot {\bm v}^{\ast },{\bm v}^{*}_{\perp} \right\} . \label{4.13}
\end{equation}
The final choice of $\Phi _{2}$ does not appear to be unique. This
is discussed further in Appendix \ref{ap2}. For the purposes of
the next section, it suffices to make the choice
\begin{equation}
\left\{ \Phi _{i}\right\} =\left\{
1,\frac{v^{*2}}{d}+\frac{1}{2},\widehat{\bm k} \cdot {\bm v}^{\ast
},{\bm v}^{*}_{\perp} \right\} . \label{4.14}
\end{equation}
The function $\Phi _{2}$ is not an eigenfunction of the adjoint
operator $\mathcal{L}^{* \dagger}$, but it is easily verified that
the biorthogonality conditions (\ref{4.11}) are satisfied.

\section{Navier-Stokes Order Modes}
\label{s5}

In the previous section, the hydrodynamic modes were identified in
the long wavelength limit. Assuming analyticity, their existence
at finite wavevectors can be inferred. Furthermore, their explicit
construction is possible by perturbation theory. This construction
provides a critical test of the internal consistency of other
quite different approaches (e.g., the Chapman-Enskog method
discussed above and in Appendix \ref{ap1}). In particular, the
detailed form of the eigenvalues and the dependence of the
associated transport coefficients on the restitution coefficient
should be exactly the same. This is demonstrated to Navier-Stokes
order in this section.

Return now to the eigenvalue problem (\ref{4.1}) and consider the
case of $ k<<1$. Look for solutions with the expansion  (a more
complete characterization of the conditions for this perturbation
expansion is given below),
\begin{equation}
\Psi _{i}(\mathbf{k})=\Psi _{i}(0)+k\Psi _{i}^{(1)}+k^{2}\Psi
_{i}^{(2)}+ \ldots, \label{5.1}
\end{equation}
\begin{equation}
\lambda _{i}(\mathbf{k})=\lambda _{i}(0)+k\lambda
_{i}^{(1)}+k^{2}\lambda _{i}^{(2)}+ \ldots . \label{5.2}
\end{equation}
The reference eigenfunctions $\Psi _{i}(0)$ and eigenvalues
$\lambda _{i}(0)$ are taken to be the long wavelength hydrodynamic
results of Eqs.\ (\ref{4.9}) and (\ref{4.10}). As already
indicated, there is a $d$-fold degeneracy for the eigenvalue
$\lambda(0)=-\zeta ^{\ast}_{0}/2$. However, the $d$-dimensional
subspace spanned by its eigenvectors was naturally partitioned by
symmetry into the longitudinal and transverse components. The
eigenvectors for the transverse modes decouple from the remaining
three modes even at finite wavevector for the same symmetry
reasons, so there are no complications of degenerate perturbation
theory. In the longitudinal subspace, all the eigenvalues are
distinct, {\em except in the elastic limit} $\alpha =1$, where
there is a three-fold degeneracy in this subspace. Thus the two
cases of unperturbed reference states with $\alpha =1$ and $\alpha
<1$ must be distinguished. In the former case, the eigenvalues
behave as
\begin{equation}
\lambda _{i}({\bm k},\alpha )=\left[ \lambda _{i}(0)+k\lambda
_{i}^{(1)}+k^{2}\lambda _{i}^{(2)}+ \dots \right]_{\alpha
=1}+\text{terms of order } (1-\alpha ),  \label{5.3}
\end{equation}
and are regular in $(1-\alpha )$. This occurs when the degree of
dissipation because of inelasticity is small relative to the
effects of the spatial variation. The corresponding eigenvalues
are then similar to those of a normal gas, with $d-1$ shear
diffusion modes, two sound modes, and a heat diffusion mode. Here
attention is restricted to the more interesting and relevant
second case of fixed $ \alpha <1$ with small spatial
perturbations. It will be seen that the modes are now
qualitatively different, consistently with the results reported in
Sec. \ref{s2}, since the degeneracy of the elastic limit is lifted
at the outset by the finite dissipation.

To set up the perturbation expansion, projection operators
$\mathcal{P}_{i}$ for the biorthogonal set $\left\{ \Phi _{i},\Psi
_{i}(0)\right\} $ are defined by
\begin{equation}
\mathcal{P}_{i}X=\Psi _{i}(0)\left( \Phi _{i},X\right) ,
\label{5.6}
\end{equation}
for an arbitrary element $X$ in the Hilbert space. The eigenvalue
problem (\ref{4.1}) can be rearranged as
\begin{equation}
\left[ \mathcal{L}^{\ast }-\lambda _{i}\left( 0 \right) \right]
\Psi _{i}\left( {\bm k}\right) =\left[ \lambda _{i}\left( k\right)
-\lambda _{i}\left( 0\right) -i {\bm k} \cdot {\bm v}^{\ast
}\right] \Psi _{i}\left({\bm k} \right) .  \label{5.7}
\end{equation}
Then, operating on both sides of this equation with
$\mathcal{Q}_{i} \equiv 1 - \mathcal{P}_{i}$, and using the
property $\left[ \mathcal{L}^{\ast }-\lambda _{i}\left( 0\right)
\right] \mathcal{P}_{i}=0$, this becomes
\begin{equation}
\mathcal{Q}_{i} \left[ \mathcal{L}^{\ast }-\lambda _{i}\left(
0\right) \right] \mathcal{Q}_{i} \Psi _{i}\left({\bm k}\right)
=\mathcal{Q}_{i} \left[ \lambda _{i}\left( k\right) -\lambda
_{i}\left( 0\right) -i {\bm k} \cdot {\bm v}^{\ast }\right] \Psi
_{i}\left( {\bm k}\right) .  \label{5.8}
\end{equation}
By construction, the right side of the above equation is
orthogonal to the null space for the adjoint of $\left[
\mathcal{L}^{\ast }-\lambda _{i}\left( 0\right) \right] $ , and
the Fredholm alternative assures solutions to this equation \cite
{Fred}. The eigenvalue problem for $\Psi _{i}\left({\bm k}\right)
$ is determined only up to an overall scale factor, amounting to
the choice of normalization. It is convenient to choose
\begin{equation}
\left( \Phi _{i},\Psi _{i}\left( \mathbf{k}\right) \right) = 1,
\label{5.8a}
\end{equation}
implying $\left(\Phi_{i},\Psi_{i}^{(n)} \right)=0$ for $n \geq 1$.
With this, Eq.\ (\ref{5.8}) gives two sets of equations for the
eigenvalues and eigenvectors,
\begin{equation}
\lambda _{i}\left({\bm k}\right) =\lambda _{i}\left( 0\right)
+\biggl( \Phi _{i},i {\bm k} \cdot {\bm v}^{\ast }\Psi _{i}\left(
\mathbf{k}\right) \biggr) +\biggl( \Phi _{i},\left[
\mathcal{L}^{\ast }-\lambda _{i}\left( 0\right) \right]
\mathcal{Q}_{i}\Psi _{i}\left( \mathbf{k}\right) \biggr),
\label{5.9}
\end{equation}
\begin{equation}
\mathcal{Q}_{i}\Psi _{i}\left({\bm k}\right) =\left\{
\mathcal{Q}_{i}\left[ \mathcal{L}^{\ast }-\lambda _{i}\left(
\mathbf{0}\right) \right] \mathcal{Q}_{i} \right\}
^{-1}\mathcal{Q}_{i}\left[ \lambda _{i}\left( k\right) -\lambda
_{i}\left( 0 \right) -i {\bm k} \cdot {\bm v}^{\ast }\right] \Psi
_{i}\left({\bm k}\right).  \label{5.10}
\end{equation}

To zeroth order in $k$, these equations give the hydrodynamic
modes of the last section in the long wavelength limit,
consistently. To first order in $k$, the eigenvectors are
\begin{eqnarray}
\Psi _{i}^{(1)}({\bm k}) &=&\mathcal{Q}_{i}\Psi _{i}^{(1)}({\bm
k})= \left\{ \mathcal{Q}_{i} \left[ \mathcal{L} ^{\ast }-\lambda
_{i}\left( 0\right) \right] \right\}^{-1} \mathcal{Q}_{i} \left(
\lambda _{i}^{(1)}-i\widehat{\bm k} \cdot {\bm v}^{\ast }\right)
\Psi
_{i}\left(0\right)  \nonumber \\
&=&- \left\{ \mathcal{Q}_{i} \left[ \mathcal{L}^{\ast }-\lambda
_{i}\left( 0 \right) \right]
\right\}^{-1}\mathcal{Q}_{i}i\widehat{\bm k}\cdot {\bm v}^{\ast
}\Psi _{i}\left( 0\right).  \label{5.11}
\end{eqnarray}
The first equality in the above transformations is a consequence
of the normalization condition (\ref{5.8a}). For the first order
eigenvalues it is found:
\begin{equation}
\lambda _{i}^{(1)}=\biggl( \Phi _{i},i\widehat{\bm k}\cdot {\bm
v}^{\ast } \Psi _{i}\left( 0\right) \biggl) +\left( \Phi
_{i},\mathcal{ L}^{\ast } \mathcal{Q}_{i} \Psi _{i}^{(1)}\right)
=0.
\end{equation}
Now the last equality follows from the fact that each $\Phi _{i}$
and $\Psi _{i}\left(0\right) $ have the same definite parity under
the change ${\bm v}^{\ast }\rightarrow - {\bm v}^{\ast }$ and the
distribution function of the HCS,  $\phi(v^{*})$, defining the
scalar product is invariant under this change. Thus the first term
on the right side vanishes. The second term also vanishes for
similar reasons, since $\Phi _{i}$ and $\Psi _{i}^{(1)}$ have
opposite parity and $ \mathcal{L}^{\ast }$ is invariant under the
change in sign of the velocity.

To second order in $k$ the eigenvectors and eigenvalues are given
by
\begin{equation}
\mathcal{Q}_{i}\Psi _{i}^{(2)}=-\left\{ \mathcal{Q}_{i}\left[
\mathcal{L}^{\ast }-\lambda _{i}\left(0\right) \right]
\right\}^{-1} \mathcal{Q}_{i} i\widehat{\bm k} \cdot {\bm v}^{\ast
}\Psi _{i}^{(1)}, \label{5.11a}
\end{equation}
\begin{equation}
\lambda _{i}^{(2)}=\left( \Phi _{i},i {\bm k} \cdot {\bm v}^{\ast
}\Psi _{i}^{(1)}\right) +\left( \Phi _{i},\mathcal{L}^{\ast
}\mathcal{Q}_{i}\Psi _{i}^{(2)}\right).   \label{5.12}
\end{equation}
The above expressions for the eigenvalues are evaluated in
Appendix \ref{ap3}. The results have the same forms as given in
Eqs.\ (\ref{2.23}). Furthermore, the reduced transport
coefficients $\eta ^{\ast },$ $\kappa^{*} $, and $\zeta _{1}^{\ast
}$ are determined from the same integral equations as following
from the Chapman-Enskog solution summarized in Appendix \ref{ap1}.
This confirms that the hydrodynamic modes determined from the
spectrum of the linearized Boltzmann equation are consistently
determined to Navier-Stokes order by both methods.

\section{``Ageing'' to Hydrodynamics}
\label{s6}

The existence of hydrodynamic excitations only assures that there
is a hydrodynamic contribution to the dynamics of small
perturbations of the HCS. To establish a description in terms of
these hydrodynamic excitations alone, it is necessary to
characterize the rest of the excitations in the spectrum. For
gases with elastic collisions, it has been shown that for
sufficiently small $k$ the hydrodynamic excitations are smaller in
magnitude than all other excitations, and bounded away from them
\cite{Mc89,Scharf}. Consequently, there is a time scale beyond
which only the hydrodynamic excitations persist. It is on this
space and time scales that hydrodynamics in the usual sense
applies. Typically, the conditions are wavelengths larger than the
mean free path and times later than the mean free time. This
leaves a large domain of macroscopic space and time scales for
hydrodynamics.

The extension of this concept of ``ageing to hydrodynamics'' for
granular gases is expected, but its verification is not so
straightforward. The mathematical analysis for elastic collisions
does not transfer to the granular gas due to the significant
differences in the linear collision operator. There are
qualitative differences in the hydrodynamic modes. For example,
the fact that energy is not conserved means that the hydrodynamic
excitations cannot be made arbitrarily small simply by making $ k
$ small. The fastest decaying hydrodynamic modes is that with
eigenvalue $\zeta _{0}^{\ast }/2$ at long wavelengths. This has
its maximum value at large dissipation, and the question arises as
to whether the time scale for this mode can become comparable to
or exceed those of the non-hydrodynamic modes at strong
dissipation.

\subsection{Model Kinetic Equation}

Current analysis of the spectrum of the linearized Boltzmann
operator for granular gases appears to be limited to the
hydrodynamic excitations discussed here, with no characterization
of the rest of the spectrum as yet. Consequently, in the remainder
of this presentation these questions are addressed in the context
of a model kinetic equation . This model \cite{BMyD96} is an
extension of the familiar Bhatnager, Gross, Krook (BGK) single
relaxation time model for normal gases \cite{Ce75}. The Boltzmann
equation can be formally written in the form
\begin{equation}
\left( \frac{\partial }{\partial t}+\mathbf{v}\cdot \nabla \right)
f=-\nu \left( f-g\right) .  \label{6.1}
\end{equation}
There are two significant differences of the model considered here
with respect to the original Boltzmann equation. First, the
collision frequency is replaced by a velocity independent function
of the local density and temperature, $\nu =\nu (n,T)$. Second,
the gain term of the Boltzmann equation is replaced by $\nu g$,
where $ g$ is taken to be a Gaussian function of the velocity,
\begin{equation}
g({\bm r},{\bm v},t)=n\left[ \frac{b(T)}{\pi }\right]
^{d/2}e^{-b(T)V^{2}}. \label{6.2}
\end{equation}
Here ${\bm V}$ is the peculiar velocity defined bellow Eq.\
(\ref{2.10}). The parameter $b$ is chosen to be function of the
local density and temperature so as to enforce the moment
conditions (\ref{2.5}) above. This leads to the identification
\begin{equation}
b(T)=\frac{m}{2T\left( 1-\zeta / \nu \right) },  \label{6.3}
\end{equation}
In this way, it is assured that the exact macroscopic balance
equations (\ref{2.6})-(\ref{2.10}) are preserved by the model.
Then the Chapman-Enskog method leads to the same Navier-Stokes
hydrodynamic equations as for the Boltzmann equation, with only
the transport coefficients being different. In the following, it
will be considered that the expression for the collision frequency
$\nu$ is chosen as scaling with $nT^{1/2}$, in order to mimic the
hard sphere behavior. Dimensional analysis then implies the same
scaling for the cooling rate $\zeta$. Moreover, note that
consistency of the model kinetic equation requires that $\zeta
<\nu$ for all values of $\alpha$. A possible choice for the
cooling rate is to be the same as obtained from the Boltzmann
equation by using a local Maxwellian for the distribution
function. In the same spirit, the collision frequency can be fixed
by fitting one of the transport coefficients of the model to that
obtained from the Boltzamnn equation by the Chapman-Enskog
procedure in the first Sonine approximation. If the shear
viscosity $\eta$ is used, the above choices lead to
\cite{Baskaran04}
\begin{equation}
\nu ({\bm r},t)= \frac{(3-3\alpha+2d)(1+\alpha)}{4d}\,
\nu_{0}({\bm r},t), \label{6.4}
\end{equation}
\begin{equation}
\zeta({\bm r},t)= \frac{(2+d)(1-\alpha^{2})}{4d}\, \nu_{0}({\bm
r},t), \label{6.4a}
\end{equation}
where $\nu _{0}({\bm r},t)$ is an average local collision
frequency,
\begin{equation}
\nu _{0}({\bm r},t)=\frac{8 \pi^{(d-1)/2} n \sigma^{d-1}}{(2+d)
\Gamma (d/2)}\ \left( \frac{T}{m} \right)^{1/2}. \label{6.5}
\end{equation}
The above expressions yield
\begin{equation}
\frac{\zeta({\bm r},t) }{\nu ({\bm r},t)
}=\frac{(2+d)(1-\alpha)}{3-3\alpha +2d } \leq \frac{2+d}{3+2d}<1,
\label{6.6}
\end{equation}
in agreement with the model consistency requirement.

This kinetic model reduces to the BGK model for normal gases at
$\alpha =1$ \cite{Ce75}. Otherwise it reproduces all of the
qualitative features of the granular Boltzmann equation, including
a nontrivial HCS and the same hydrodynamic excitations discussed
in the sections above. In some respects, the model kinetic
equation is more complex than the Boltzmann equation since the
collision operator is a nonlinear functional of $f$ \ through the
dependence of $g$ on $T$ and $\mathbf{u}$. However, the linearized
model kinetic equation for small perturbations of the HCS is
considerably simpler than that for the Boltzmann equation, as
shown below.

\subsection{Model HCS and Linear Model Kinetic Equation}

The HCS equation (\ref{3.3}) for this model becomes in
dimensionless form
\begin{equation}
\frac{\zeta^{*}_{0}}{2} \frac{\partial}{\partial {\bm v}^{*}}
\cdot \left( {\bm v}^{*} \phi \right) +\nu _{0}^{\ast }\phi =\nu
_{0}^{\ast }\left[ \pi \left( 1-\zeta^{*}_{0}/ \nu^{*}_{0} \right)
\right] ^{-d/2} \exp \left( -
\frac{v^{*2}}{1-\zeta_{0}^{*}/\nu_{0}^{*}} \right), \label{6.7}
\end{equation}
where $\zeta^{*}_{0} \equiv \ell \zeta_{hcs}/v_{hcs}(t)$, as in
the preceding sections and, consistently, $\nu^{*}_{0} \equiv \ell
\nu_{hcs}/v_{hcs}(t)$. The solution to the above equation is
\begin{equation}
\phi (v^{\ast })=\nu ^{\ast }_{0} \left[ \pi \left(
1-\zeta^{\ast}_{0} / \nu ^{\ast }_{0} \right) \right]
^{-d/2}\int_{0}^{\infty }ds\, \exp \left[ -\left( \frac{d
\zeta^{*}_{0}}{2} +\nu ^{\ast }_{0} \right) s \right]\, \exp
\left( -\frac{e^{-\zeta_{0}^{\ast }s}v^{\ast 2}}{1-\zeta ^{\ast
}_{0}/ \nu ^{\ast }_{0}}\right). \label{6.8}
\end{equation}
It is easily verified that this distribution exhibits algebraic
decay for large velocities,
\begin{equation}
\phi (v^{\ast }) \sim \frac{p \pi ^{-d/2}}{2} \left(
\frac{p}{p-2}\right) ^{-p/2}\Gamma \left( \frac{p+d}{2}\right)
v^{\ast -\left(p+d \right) },   \label{6.9}
\end{equation}
with $ p = 2\nu ^{\ast }_{0} / \zeta ^{\ast }_{0}$. Therefore,
moments of degree $p$ or greater do not exist.

The linearized kinetic model equation for small perturbations of
the HCS is obtained in Appendix \ref{ap4}, with the result, in the
dimensionless variables of the previous sections,
\begin{equation}
\left( \partial _{s}+i {\bm k} \cdot {\bm v}^{\ast
}\mathbf{+}\mathcal{L}^{\ast }_{m} \right) \widetilde{\Delta}
({\bm k}, {\bm v}^{\ast }, s)=0. \label{6.10}
\end{equation}
The linear collision operator in this case is
\begin{equation}
\mathcal{L}^{\ast }_{m}=\sum_{i}\lambda
_{i}(0)\mathcal{P}_{i}+\mathcal{Q}\left( \nu _{hcs}^{\ast
}+\frac{\zeta_{0}^{*}}{2} \phi ^{-1} \frac{\partial}{\partial {\bm
v}^{*}} \cdot {\bm v}^{*} \phi \right) \mathcal{Q}. \label{6.11}
\end{equation}
The projection operators $\mathcal{P}_{i}$ are the same as defined
in Eq.\ (\ref{5.6}) and
\begin{equation}
\mathcal{P} \equiv \sum_{i}\mathcal{P}_{i},\quad \mathcal{Q}
\equiv 1-\mathcal{P}. \label{6.12}
\end{equation}
In the above expressions, the summations are over the hydrodynamic
modes. The first term on the right hand side of Eq.\, (\ref{6.11})
is the projection onto the hydrodynamic eigenfunctions, while the
second one is orthogonal to this subspace. Consequently, the
spectrum of $\mathcal{L}^{\ast }_{m}$ for the model kinetic
equation has the same $k=0$ hydrodynamic eigenfunctions and
eigenvalues as the Boltzmann equation, i.e. it is
\begin{equation}
\mathcal{L}^{\ast }_{m} \Psi _{i} (0)=\lambda _{i}\left( 0\right)
\Psi _{i}(0) , \quad i=1, \ldots, d+2.  \label{6.13}
\end{equation}

Furthermore, the structure of $\mathcal{L}^{\ast }_{m}$ decomposes
into operators defined in the hydrodynamic subspace and its
orthogonal complement. This allows more detailed analysis of the
non-hydrodynamic spectrum. The associated eigenfunctions lie in
the orthogonal complement $ \mathcal{Q}$. Consider the general
form for $\Psi_{Q}=\mathcal{Q} \Psi$
\begin{eqnarray}
\biggl( \Psi_{Q},[ \mathcal{L}^{\ast }-\sum_{i}\lambda
_{i}(0)\mathcal{P }_{i} ] \Psi_{Q} \biggr)  &=&\biggl(
\Psi_{Q},\left( \nu _{0}^{\ast }+ \frac{\zeta_{0}^{*}}{2} \phi
^{-1} \frac{\partial}{\partial {\bm v}^{*}} \cdot
{\bm v}^{\ast }\phi \right) \Psi_{Q} \biggr)   \nonumber \\
&=&\biggl( \left( \nu _{0}^{\ast }-\frac{\zeta_{0}^{*}}{2} {\bm v}
^{\ast }\cdot \frac{\partial}{\partial {\bm v}^{*}} \right)
\Psi_{Q}, \Psi_{Q} \biggr). \label{6.15}
\end{eqnarray}
The second term in the scalar product above can be simplified as
\begin{eqnarray}
\frac{\zeta_{0}^{*}}{2} \biggl( {\bm v}^{\ast } \cdot
\frac{\partial}{\partial {\bm v}^{*}} \Psi_{Q}, \Psi_{Q} \biggr)
&=&\frac{\zeta_{0}^{*}}{4 }\biggl( 1,{\bm v}^{\ast } \cdot
\frac{\partial}{\partial {\bm v}^{*}} \Psi_{Q}^{2} \biggr)
 \nonumber \\
&=&-\frac{\zeta_{0}^{*}}{4} \biggl( \phi ^{-1}
\frac{\partial}{\partial {\bm v}^{*}} \cdot \left( {\bm v}^{\ast
}\phi \right) ,\Psi_{Q}^{2} \biggr)   \nonumber \\
&=&\frac{\nu_{0}^{*}}{2} \left( \Psi_{Q},\Psi_{Q} \right)
-\frac{\nu^{*}_{0}}{2} \left( \phi ^{-1}g_{0}^{\ast
},\Psi_{Q}^{2}\right), \label{6.16}
\end{eqnarray}
where
\begin{equation}
g_{0}^{*}(v^{*})=n_{hcs}^{-1} v^{d}(t)g_{hcs}(v,t) \geq 0.
\label{6.16a}
\end{equation}
Then Eq.\, (\ref{6.15}) gives the desired inequality
\begin{equation}
\frac{\left( \Psi_{Q },\left[ \mathcal{L}^{\ast
}_{m}-\sum_{i}\lambda _{i}(0) \mathcal{P}_{i}\right]
\Psi_{Q}\right)}{\left(\Psi_{Q}, \Psi_{Q} \right) }
=\frac{\nu_{0}^{*}}{2} \left[ 1+\frac{\left( \Psi_{Q},\phi
^{-1}g^{\ast }_{0} \Psi_{Q} \right)
}{\left(\Psi_{Q},\Psi_{Q}\right) }\right] >
\frac{\nu_{0}^{*}}{2}\, . \label{6.17}
\end{equation}
Therefore, the non-hydrodynamic spectrum of $\mathcal{L}^{\ast
}_{m}$ consists of points (or continuum) with real parts larger
than $\nu _{0}^{\ast }/2$. The fastest decaying hydrodynamic
eigenvalue is at $\zeta _{0}^{\ast }/2$. The hydrodynamic
excitations are isolated from the rest of the spectrum for $ \zeta
_{0}^{\ast }<\nu _{0}^{\ast }$, as it is always the case. Assuming
analyticity in $k$, the hydrodynamic modes at finite wavelength
also will be isolated from the rest of the spectrum for
sufficiently small $k$.

In summary, the kinetic model considered here illustrates the
expected behavior for the Boltzmann equation. The linearized
kinetic equation for small perturbations of the HCS characterizes
the complete complex response. Among the excitations there are
$d+2$ hydrodynamic modes. At long wavelengths, these modes are
isolated from and have smaller eigenvalues than the rest of the
spectrum. Hence, there is a sufficiently long time scale on which
only the hydrodynamic excitations persist.

This kinetic model  can be used to explore the role of
hydrodynamics in great detail as the linear equation can be solved
exactly \cite{Baskaran04}. For example, it has been shown that the
hydrodynamic modes extend to very short wavelengths far beyond the
validity of the Navier-Stokes approximation.

\section{Discussion}

The objective here has been to explore the role of hydrodynamics
for a granular gas by a direct analysis of the spectrum of the
linear inelastic Boltzmann equation. This analysis has been shown
to lead to results equivalent to those obtained previously based
on the Chapman-Enskog method to solve the Boltzmann equation.
However, the current method is more straightforward and less
susceptible to subjective questions about the applicability of the
Chapman-Enskog method to granular gases. In addition, the
formulation of the problem in terms of the spectrum of the
linearized collision operator is the proper setting to explore the
context for dominance of a hydrodynamic description.

The eigenvalue problem posed in Sec. \ref{s4}, together with the
macroscopic balance equations leads to a precise definition of the
hydrodynamic modes. Here, it has been established that they exist
for sufficiently long wavelengths. It remains to explore the form
and extent to which they are meaningful at shorter wavelengths,
and other methods to solve the eigenvalue problem are available to
complement the simple perturbation theory applied here. This has
been done in ref. \cite{Baskaran04} for the kinetic model
introduced in  Sec. \ref{s5}, demonstrating the extension of the
hydrodynamic modes to wavelengths an order of magnitude shorter
than those required for the Navier-Stokes approximation .

A second relevant result of the analysis here is the
identification of the hydrodynamic eigenfunctions. This allows
calculation of the hydrodynamic component for properties of
interest, and provides the means to explore dynamical mechanisms
beyond the Boltzmann description based on hydrodynamics (e.g.,
mode coupling phenomena for fluctuations). For normal gases, these
eigenfunctions are the summational invariants $(1,v^2,{\bm v})$ in
the long wavelength limit. For a granular gas, they are replaced
by derivatives of the logarithm of the HCS distribution function,
which behave quite differently for large velocities, as
illustrated in Figs.\ \ref{fig1} and \ref{fig2}.

The hydrodynamic excitations are usually interpreted as a set of
$d+2$ modes associated with the density, temperature, and flow
field of the macroscopic balance equations. Here, they are simply
$d+2$ points associated with the spectrum of the linearized
inelastic Boltzmann equation for response to small perturbations.
It is often claimed that the temperature should not be included in
the set of hydrodynamic fields, as the energy is not conserved.
However, the analysis here shows that in the proper reduced
variables all but one eigenvalue is non-zero in the long
wavelength limit. Thus, the exclusion of one field from the
description will not recover a smaller set of eigenvalues
clustered around zero at long wavelengths. Instead, the $d+2$
modes have the following qualitative behavior. They are all
clustered near zero for weak dissipation, but become increasingly
separated at stronger dissipation. This means that within the
hydrodynamic description there are time scales set by both the
wavelength and the cooling rate and these can be quite different.

A second question is the isolation of these  hydrodynamic modes
from the rest of the spectrum. At weak dissipation, it is expected
that this is the case since the hydrodynamic eigenvalues are all
small and the non-hydrodynamic spectrum is expected to be of the
order of the collision frequency. At strong dissipation, there is
a hydrodynamic mode of the order of $ \zeta _{0}^{\ast }/2$, which
can be of the order of the collision frequency. It remains an open
question regarding the size of this eigenvalue relative to the
non-hydrodynamic spectrum. It is possible that the isolation of
the hydrodynamic spectrum places some restriction on the degree of
dissipation. However, the analysis based on the kinetic model of
Sec. \ref{s6} suggests this may not be the case. For the kinetic
model, all of the hydrodynamic spectra remain isolated from all of
the non-hydrodynamic spectra for any degree of dissipation. In
this case, the $d+2$ hydrodynamic modes define a dominant set on a
sufficiently large time scale, even when the separation of times
among the hydrodynamic modes is significant.

 The idealized model for a granular gas discussed here can be
made more realistic by considering a more complex binary collision
rule. In particular, a velocity dependent restitution coefficient
(which approaches unity as the relative velocity goes to zero) and
tangential friction are two additional qualitative features of
real granular fluids. However, the general form of the
hydrodynamic equations (macroscopic balance equations) is
unchanged in that case; only the detailed values of the transport
coefficients, pressure, and cooling rates differ from those of the
present model. Similarly, the collision operator of the Boltzmann
equation becomes more complex but its properties relevant for the
macroscopic balance equations are the same. Most of the analysis
given here depends only on those properties rather than the
detailed form of the collision operator. Consequently, it is
expected that the implications of our simple model extend to more
realistic models as well.

The analysis of the linear Boltzmann collision operator for
normal, elastic gases is quite complete \cite{Mc89}. It is hoped
that the beginning provided here for granular gases will provoke
the new mathematical analysis required for this case as well.

\section{Acknowledgements}
J.W.D thanks A. Baskaran for helpful discussions and both authors
thank M.J. Ruiz-Montero for providing Figs. \ref{fig1} and
\ref{fig2}. The research of J.W.D. was supported in part by
Department of Energy Grant DE-FG02ER54677. The research of J.J.B.
was supported by the Ministerio de Educaci\'{o}n y Ciencia (Spain)
through Grant No. BFM2005-01398 (partially financed by FEDER
funds).

\appendix

\section{Chapman-Enskog Results}
\label{ap1}

The Chapman-Enskog method constructs a solution to the Boltzmann
equation whose space and time dependence occurs entirely through
the hydrodynamic fields and their gradients. For small spatial
gradients, i.e. small relative variation of the hydrodynamic
fields over a mean free path, the solution reads (we use the same
notation as in ref.\ \cite{BDKyS98})
\begin{equation}
f({\bm r},{\bm v},t)=f_{hcs}^{(0)}({\bm r},{\bm v},t) +
\mathcal{A} \cdot {\bm \nabla} \ln T + \mathcal{B} \cdot {\bm
\nabla} \ln n + {\sf C} : {\bm \nabla} {\bm u}+ {\cal M}
\nabla^{2} T + {\cal N} \nabla^{2} n. \label{a.1}
\end{equation}
Consider the contributions to the heat flux and the momentum flux
from the terms of first order in the gradients of the above
expression. Since these fluxes appear in the macroscopic balance
equations under a gradient, the resulting contributions are of
second order in the gradients (Navier-Stokes order). For
consistency, the cooling rate which does not occur under a
gradient in the balance equations must be calculated to second
order. Therefore, the last two terms on the right hand side of
Eq.\ (\ref{a.1}) give contributions to the cooling rate at this
order, but lead to higher order (Burnett) terms in the heat and
momentum fluxes. Additional non-linear terms of second order in
the gradients coming from the cooling rate have been omitted in
Eq.\ (\ref{a.1}) as only the linear hydrodynamic equations are
considered here. The reference distribution function
$f^{(0)}_{hcs}({\bm r},{\bm v},t)$ has the same functional form as
the distribution function of the HCS discussed at the beginning of
Sec.\ \ref{s3}, but scaled with respect to the local exact
hydrodynamic fields at time $t$ for the generally non-uniform,
non-stationary state. Therefore, it can be considered as the {\em
local} HCS distribution function.

The dissipative part of the pressure tensor and the heat flux are
given by Eqs.\ (\ref{2.11}) and (\ref{2.12}), respectively. The
transport coefficients in these equations are determined from the
functions $\mathcal{A}$, $\mathcal{B}$, and ${\sf C}$ appearing in
Eq.\ (\ref{a.1} ) through
\begin{equation}
\eta = - \frac{1}{d^{2}+d-2}\, \int d{\bm v}\, {\sf D}({\bm v}) :
{\sf C}({\bm v}), \quad \kappa= - \frac{1}{dT}\,  \int d{\bm v}
 {\bm S} ({\bm v}) \cdot \mathcal{A}({\bm v}), \nonumber
\end{equation}
\begin{equation}
\mu =- \frac{1}{dn}\, \int d{\bm v} {\bm S}({\bm v}) \cdot
\mathcal{B}({\bm v}), \label{a.3}
\end{equation}
where the functions ${\sf D}({\bm v})$ and ${\bm S}({\bm v})$ are
defined by
\begin{equation}
{\sf D} \equiv m \left( {\bm v} {\bm v} - \frac{v^{2}}{d}  {\sf I}
\right), \quad {\bm S}= \left( \frac{m v^{2}}{2} -\frac{d+2}{2}\,
T \right) {\bm v}. \label{a.4}
\end{equation}

The functions $\mathcal{A}({\bm v},n,T)$, $\mathcal{B}({\bm
v},n,T)$, and ${\sf C}({\bm v},n,t)$ are solutions of the integral
equations
\begin{equation}
\left( -\zeta_{hcs}^{(0)} T \frac{\partial}{\partial T} + L^{(0)}
- \frac{\zeta^{(0)}_{hcs}}{2} \right) \mathcal{A} ={\bm A},
\label{a.5}
\end{equation}
\begin{equation}
\left( -\zeta_{hcs}^{(0)} T \frac{\partial}{\partial T} + L^{(0)}
\right) \mathcal{B} ={\bm B}+ \zeta_{hcs}^{(0)} \mathcal{A},
\label{a.6}
\end{equation}
\begin{equation}
\left( -\zeta_{hcs}^{(0)} T \frac{\partial}{\partial T} + L^{(0)}
\right) {\sf C} ={\sf  G}, \label{a.7}
\end{equation}
with the definitions
\begin{equation}
L^{(0)}(t)X({\bm v})=-J[f_{hcs}^{(0)},X]-J[X,f_{hcs}^{(0)}],
\label{a.8}
\end{equation}
\begin{equation}
{\bm A}=\frac{\bm v}{2} \frac{\partial}{\partial {\bm v}} \cdot
\left( {\bm v} f_{hcs}^{(0)} \right) -\frac{T}{m} \frac{\partial
f_{hcs}^{(0)}}{\partial {\bm v}}, \quad {\bm B}= - {\bm v}
f_{hcs}^{(0)}-\frac{T}{m} \frac{\partial f_{hcs}^{(0)}}{\partial
{\bm v}}, \nonumber
\end{equation}
\begin{equation}
{\sf G}=\frac{\partial}{\partial {\bm v}} \left( {\bm v}
f_{hcs}^{(0)} \right)- \frac{\sf I}{d} \frac{\partial}{\partial
{\bm v}} \cdot \left( {\bm v} f_{hcs}^{(0)} \right). \label{a.9}
\end{equation}
Moreover, $\zeta_{hcs}^{(0)}$ is the local form of $\zeta_{hcs}$.

The transport coefficients arising from the cooling rate are best
described by first making the expression for the cooling rate more
explicit. In general, it is given by a bilinear functional of the
distribution function,
\begin{equation}
\zeta =\zeta \left[ f,f\right],  \label{a.9a}
\end{equation}
where
\begin{eqnarray}
\zeta \left[ X,Y\right] &=&-\frac{m}{dnT}\int d{\bm v}\,
 v^{2}J[X,Y]  \nonumber \\
&=&\left( 1-\alpha ^{2}\right) \frac{m \pi ^{\left( d-1\right)
/2}\sigma ^{d-1}}{4 d n T \Gamma \left( \frac{d+3}{2}\right) }\int
d {\bm v} \int d {\bm v}_{1}\, g^{3} X({\bm v}) Y ({\bm v}_{1}) =
\zeta [Y,X]. \label{a.10}
\end{eqnarray}
The linear contributions from the cooling rate to second order
order in the gradients are then
\begin{equation}
\zeta_{L}^{(2)}= \zeta_{1} \nabla^{2} T+ \zeta_{2} \nabla^{2} n,
\label{a.11}
\end{equation}
with
\begin{equation}
\zeta_{1}= 2 \zeta [ \mathcal{M}, f_{hcs}^{(0)}], \quad \zeta_{2}=
2 \zeta [ \mathcal{N}, f_{hcs}^{(0)} ]. \label{a.12}
\end{equation}
The integral equations for $\mathcal{M}$ and $\mathcal{N}$ are
found to be
\begin{equation}
\left( -\zeta_{hcs}^{(0)} T \frac{\partial}{\partial
T}-\frac{3}{2} \zeta_{hcs}^{(0)} +L^{(0)} \right) \mathcal{M}= T
\zeta_{1} \frac{\partial f_{hcs}^{(0)}}{\partial T}- \left(
\frac{2 \kappa}{dn} \frac{ \partial f^{(0)}_{hcs}}{\partial T} +
\frac{1}{dT} \mathcal{A} \cdot {\bm v} \right), \label{a.13}
\end{equation}
\begin{eqnarray}
\left( -\zeta_{hcs}^{(0)} T \frac{\partial}{\partial T} +L^{(0)}
\right) \mathcal{N} & = & T \zeta_{2} \frac{\partial
f_{hcs}^{(0)}}{\partial T} + \frac{T \zeta_{hcs}^{(0)}}{n}\
\mathcal{M} \nonumber \\
& & -\left( \frac{2 \mu}{dn} \frac{\partial
f_{hcs}^{(0)}}{\partial T}+\frac{1}{dn} \mathcal{B} \cdot {\bm v}
\right). \label{a.14}
\end{eqnarray}

The above results can be easily transformed to be expressed in
terms of the reduced units and quantities introduced in Secs.
\ref{s2} and \ref{s3}. Dimensionless transport coefficients are
again defined by Eqs.\ (\ref{2.19b}) and (\ref{2.19c}), but
replacing $n_{hcs}$ and $T_{hcs}$ by their local values $n$ and
$T$. Of course, now it is $\ell= (n \sigma^{d-1})^{-1}$ and
$v_{hcs}(t)$ is replaced by $v(t)=(2T/m)^{1/2}$. In this way, it
is obtained:
\begin{equation}
\eta^{*}=\frac{1}{d^{2}+d-2} \sum_{i,j} \left( D^{*}_{ij},
C^{*}_{ij} \right),  \quad  \kappa^{*}= \frac{1}{d^{2}} \sum_{i}
\left( S^{*}_{i}, \mathcal{A}^{*}_{i} \right), \nonumber
\end{equation}
\begin{equation}
\mu^{*}=\frac{1}{d^{2}} \sum_{i} \left( S^{*}_{i},
\mathcal{B}^{*}_{i} \right), \label{a.15}
\end{equation}
with the dimensionless functions ${\sf D}^{*}$ and ${\bm S}^{*}$
given by
\begin{equation}
{\sf D}^{*}={\bm v}^{*}{\bm v}^{*}-\frac{v^{*2}}{d}\ {\sf I},
\quad {\bm S}^{*}= \left( v^{*2}- \frac{d+2}{2} \right){\bm
v}^{*}. \label{a.16}
\end{equation}
Moreover, $\mathcal{A}^{*}$, $\mathcal{B}^{*}$, and ${\sf C}^{*}$
are now defined through the equations
\begin{equation}
\left( \mathcal{L}^{*}-\frac{\zeta_{0}^{*}}{2} \right)
\mathcal{A}^{*}={\bm A}^{*}, \label{a.17}
\end{equation}
\begin{equation}
\mathcal{L}^{*} \mathcal{B}^{*}={\bm B}^{*}+\zeta_{0}^{*}
\mathcal{A}^{*}, \label{a.19}
\end{equation}
\begin{equation}
\left( \mathcal{L}^{*} + \frac{\zeta_{0}^{*}}{2} \right) {\sf
C}^{*}= {\sf G}^{*}, \label{a.18}
\end{equation}
where the operator $\mathcal{L}^{*}$ is defined in Eq.\
(\ref{3.9}) and
\begin{equation}
{\bm A}^{*}={\bm v}^{*} \Psi_{2}(0)-{\bm \Psi}_{3}(0), \quad {\bm
B}^{*}= 2 {\bm v}^{*}-{\bm \Psi_{3}}(0), \nonumber
\end{equation}
\begin{equation}
{\sf G}^{*}={\bm v}^{*} {\bm \Psi}_{3}(0)-\frac{\sf I}{d} {\bm
v}^{*} \cdot {\bm \Psi_{3}}(0). \label{a.20}
\end{equation}
The expressions of the reduced transport coefficients
$\zeta^{*}_{1}$ and $\zeta^{*}_{2}$ are:
\begin{equation}
\zeta_{1}^{*}=\frac{1}{d} \biggl( a, \mathcal{M}^{*} \biggr),
\quad \zeta_{2}^{*}=\frac{1}{d} \biggl( a, \mathcal{N}^{*}
\biggr), \label{a.21}
\end{equation}
where
\begin{equation}
a({\bm v}^{*})= \frac{(1-\alpha^{2}) \pi^{(d-1)/2}}{2 \Gamma
\left( \frac{d+3}{2} \right)}\, \int d{\bm v}^{*}_{1}\, \phi
(v^{*}_{1}) g^{*3}. \label{a.23}
\end{equation}
The integral equations obeyed by $\mathcal{M}^{*}$ and
$\mathcal{N}^{*}$ are
\begin{equation}
\left( \mathcal{L}^{*}-\frac{\zeta_{0}^{*}}{2} \right)
\mathcal{M}^{*}= \left( \zeta^{*}_{1}-\kappa^{*} \right)
\Psi_{2}(0)+\frac{1}{d}\, \mathcal{A}^{*} \cdot {\bm v}^{*},
\label{a.24}
\end{equation}
\begin{equation}
\mathcal{L}^{*} \mathcal{N}^{*}= \zeta_{0}^{*} \mathcal{M}^{*}+
\left( \zeta_{2}^{*}-\mu^{*} \right) \Psi_{2}(0)+\frac{1}{d}\,
\mathcal{B}^{*} \cdot {\bm v}^{*}. \label{a.35}
\end{equation}

For later use, it is convenient to elaborate more the above
expression for $\zeta^{*}_{1}$. By construction, the velocity
integrals of $\mathcal{M}$ times $1$, ${\bm v}$, and $v^{2}$
vanish. This is equivalent to say that $\mathcal{M}^{*}$ is
orthogonal to the set of functions $\Phi_{i}$ defined in Eq.\
(\ref{4.14}), and in particular it verifies
$\mathcal{M}^{*}=\mathcal{Q}_{2} \mathcal{M}^{*}$, with
$\mathcal{Q}_{2}= 1-\mathcal{P}_{2}$, $\mathcal{P}_{i}$ being the
projection operator defined in Eq.\ (\ref{5.6}). Then, acting with
$\mathcal{Q}_{2}$ on both sides of Eq.\ (\ref{a.24}) it is
obtained:
\begin{equation}
\mathcal{M}^{*}=\mathcal{Q}_{2} \mathcal{M}^{*}= \frac{1}{d}\,
\left[ \mathcal{Q}_{2} \left(
\mathcal{L}^{*}-\frac{\zeta_{0}^{*}}{2} \right) \right]^{-1}
\mathcal{Q}_{2} \mathcal{A}^{*} \cdot {\bm v}^{*}. \label{a.36}
\end{equation}
For the same reason, Eq.\ (\ref{a.17}) yields
\begin{eqnarray}
\mathcal{A}^{*} & = & \mathcal{Q}_{2} \mathcal{A}^{*}=\left[
\mathcal{Q}_{2} \left( \mathcal{L}^{*}-\frac{\zeta_{0}^{*}}{2}
\right) \right]^{-1} \mathcal{Q}_{2} {\bm A}^{*} \nonumber \\
& = & \left[ \mathcal{Q}_{2} \left( \mathcal{L}^{*}
-\frac{\zeta_{0}^{*}}{2} \right) \right]^{-1} \mathcal{Q}_{2} {\bm
v}^{*} \Psi_{2}(0)-\left[ \mathcal{Q}_{2} \left( \mathcal{L}^{*}
-\frac{\zeta_{0}^{*}}{2} \right) \right]^{-1} {\bm \Psi}_{3}(0)
\nonumber \\
& = & \left[ \mathcal{Q}_{2} \left(
\mathcal{L}^{*}-\frac{\zeta_{0}^{*}}{2} \right) \right]^{-1}
\mathcal{Q}_{2} {\bm v}^{*} \Psi_{2}(0)+\frac{1}{\zeta_{0}^{*}}
{\bm \Psi}_{3}(0). \label{a.37}
\end{eqnarray}
Substitution of this expression into Eq.\ (\ref{a.36}) after some
algebra leads to
\begin{equation}
\mathcal{M}^{*}= \frac{1}{d}\ \left[ \mathcal{Q}_{2} \left(
\mathcal{L}^{*}- \frac{\zeta_{0}^{*}}{2} \right) \right]^{-1}
\mathcal{Q}_{2} {\bm v}^{*} \cdot \left[ \mathcal{Q}_{2} \left(
\mathcal{L}^{*}- \frac{\zeta_{0}^{*}}{2} \right) \right]^{-1}
\mathcal{Q}_{2} {\bm v}^{*} \Psi_{2}(0)-\frac{2}{\zeta_{0}^{*2}}
\Psi_{1}(0), \label{a.38}
\end{equation}
and use of this into Eq.\ (\ref{a.21}) gives the result
\begin{equation}
\zeta_{1}^{*}=\frac{1}{\zeta^{*}_{0}}+\frac{1}{d} \left( a,
\mathcal{M}^{*}_{1} \right), \label{a.39}
\end{equation}
with
\begin{equation}
\mathcal{M}^{*}_{1}= \frac{1}{d} \left[ \mathcal{Q}_{2} \left(
\mathcal{L}^{*}-\frac{\zeta_{0}^{*}}{2} \right) \right]^{-1}
\mathcal{Q}_{2} {\bm v}^{*} \cdot \left[ \mathcal{Q}_{2} \left(
\mathcal{L}^{*}-\frac{\zeta_{0}^{*}}{2} \right) \right]^{-1}
\mathcal{Q}_{2} {\bm v}^{*} \Psi_{2}(0). \label{a.40}
\end{equation}

Although the expression for $\zeta_{2}^{*}$ can be written in a
similar way, it will not be needed here.

\section{Adjoint Linear Operator and Biorthogonal Set}
\label{ap2} The adjoint for $\mathcal{L}^{\ast }$, $\mathcal{L}^{*
\dagger}$, is defined as usual by
\begin{equation}
\left( X,\mathcal{L}^{\ast }Y\right) =\left( \mathcal{L}^{\ast
\dagger }X,Y\right),  \label{b.1}
\end{equation}
for arbitrary $X({\bm v}^{*})$ and $Y({\bm v}^{*})$ belonging to
the Hilbert space. The explicit form of $\mathcal{L}^{*}$ is given
in Eq.\ (\ref{3.9}). From it, and using the above definition, the
expression for $\mathcal{L}^{* \dagger}$ is easily found,
\begin{eqnarray}
\mathcal{L}^{* \dagger} X ({\bm v}^{*}) &=&-\int d{\bm
v}^{*}_{1}\, \phi (v_{1}^{*}) \int d\widehat{\bm \sigma}\, \Theta
(\widehat{\bm \sigma} \cdot {\bm g}) \widehat{\bm \sigma} \cdot
{\bm g} \left[ X({\bm v}^{* \prime \prime})+X({\bm v}_{1}^{*
\prime \prime})-X({\bm v}^{*})-X({\bm
v}^{*}_{1}) \right] \nonumber \\
& & -\frac{\zeta_{0}^{*}}{2}  {\bm v}^{*} \cdot
\frac{\partial}{\partial {\bm v}^{*}} X ({\bm v}^{*}), \label{b.2}
\end{eqnarray}
where ${\bm v}^{* \prime \prime}$ and ${\bm v}_{1}^{* \prime
\prime}$ are the postcollisional velocities corresponding to ${\bm
v}^{*}$ and ${\bm v}^{*}_{1}$,
\begin{eqnarray}
{\bm v}^{* \prime \prime}& = & {\bm v}^{*}-\frac{1+\alpha}{2}\,
(\widehat{\bm \sigma} \cdot {\bm g}^{*}) \widehat{\bm \sigma},
\nonumber \\
{\bm v}^{* \prime \prime}_{1}& = & {\bm
v}^{*}_{1}+\frac{1+\alpha}{2}\, (\widehat{\bm \sigma} \cdot {\bm
g}^{*}) \widehat{\bm \sigma}. \label{b.3}
\end{eqnarray}
Equation (\ref{b.2}) gives immediately
\begin{equation}
\mathcal{L}^{\ast \dagger }1=0, \quad \quad \mathcal{L}^{\ast
\dagger } \mathbf{v}^{\ast }=-\frac{\zeta_{0}^{*}}{2} {\bm
v}^{\ast }, \label{b.4}
\end{equation}
so that $1$ and ${\bm v}^{\ast }$ are eigenfunctions of the
adjoint operator with eigenvalues $0$ and $-\zeta _{0}^{\ast }/2,$
respectively.

However, the kinetic energy is not an eigenfunction of the adjoint
operator. Direct calculation gives
\begin{equation}
\mathcal{L}^{\ast \dagger }v^{\ast 2}=-\zeta ^{\ast }_{0}v^{\ast
2}+\frac{a({\bm v}^{*})}{2},  \label{b.5}
\end{equation}
where $a({\bm v})$ is given by Eq.\ (\ref{a.23}). Nevertheless, a
biorthogonal set can be constructed from $1,$ ${\bm v}^{*}$, and
$v^{*2}$ and it is given in Eq.\ (\ref{4.14}). As noted in the
main text, the choice of this set is not unique. The conditions of
biorthogonality on $\Phi _{2}$ are
\begin{equation}
\biggl( \Phi _{2},\Psi _{2}(0) \biggr) = \left( {\bm v}^{*} \cdot
\frac{\partial \Phi_{2}}{\partial {\bm v}^{*}}, 1 \right) =1,
\label{b.6}
\end{equation}
\begin{equation}
\biggl( \Phi _{2},\Psi _{1}(0) \biggr) = \left( \Phi _{2},1\right)
-1 =0 , \label{b.7}
\end{equation}
\begin{equation}
\biggl( \Phi _{2},{\bm \Psi} _{3}(0) \biggl) = \left(
\frac{\partial \Phi_{2}}{ \partial {\bm v}^{*}}\, ,1 \right)={\bm
0} .\label{b.8}
\end{equation}
A sufficient condition to guarantee that the above relations are
verified, is that $\Phi_{2}$ have the form
\begin{equation}
\Phi _{2}({\bm v}^{*}) =A+Bb(v^{\ast }),  \label{b.9}
\end{equation}
where $b(v^{\ast })$ is an arbitrary scalar function of ${\bm
v}^{\ast }$, and  the constants $A$ and $B$ are determined from
\begin{equation}
A=1-\left( 1,b\right) B , \quad \quad  B=\left({\bm v}^{\ast }
\cdot  \frac{\partial b}{\partial {\bm v}^{*}}\, ,1 \right)^{-1}.
\label{b.10}
\end{equation}
Substitution of these expressions into Eq.\ (\ref{b.8}) yields
\begin{equation}
\Phi _{2}({\bm v}^{*})=1+ \left[ b(v^{*})- \left(1,b \right)
\right] \biggl({\bm v}^{*} \cdot \frac{\partial b}{\partial {\bm
v}^{*}}\, , 1 \biggr)^{-1}. \label{b.11}
\end{equation}
The optimal choice for $b(v^{\ast })$ would be that implying that
$\Phi_{2}({\bm v}^{*})$ is an eigenfunction of $\mathcal{L}^{*
\dagger}$ corresponding to the eigenvalue $\zeta_{0}^{*}/2$. This
is accomplished if $b(v^{*})$ is the solution to
\begin{equation}
\mathcal{L}^{\ast \dagger }b(v^{\ast })=\frac{\zeta_{0}^{*}}{2}
\left[ b(v^{\ast }) +B^{-1} \right]. \label{b.12}
\end{equation}
The solution to this equation, if it exists, has not yet been
found.

\section{Evaluation of the Perturbation Theory Results}
\label{ap3} The expansion of the hydrodynamic eigenvalues of the
linearized Boltzmann equation for small $k$ is given in Sec.
\ref{s5} with the result
\begin{equation}
\lambda _{i}(\mathbf{k})=\lambda _{i}(0)+ k^{2}\lambda _{i}^{(2)}+
\dots ,  \label{c.1}
\end{equation}
where
\begin{equation}
\left\{ \lambda _{i}(0)\right\} =\left\{
0,\frac{\zeta^{*}_{0}}{2},-\frac{\zeta^{*}_{0} }{2} \right\} ,
\label{c.2}
\end{equation}
and
\begin{eqnarray}
\lambda _{i}^{(2)} &=&\left( \Phi _{i},i\widehat{\bm k} \cdot {\bm
v}^{\ast }\Psi _{i}^{(1)}\right) +\left( \Phi
_{i},\mathcal{L}^{\ast }\mathcal{
Q}_{i}\Psi _{i}^{(2)}\right)  \nonumber \\
&=&\biggl( \Phi _{i},\widehat{\bm k} \cdot {\bm v}^{\ast} \left\{
\mathcal{Q}_{i} \left[ \mathcal{L}^{\ast }-\lambda_{i}\left(0
\right) \right] \right\}^{-1}\mathcal{Q}_{i} \widehat{\bm k} \cdot
{\bm v}^{\ast } \Psi _{i}\left( 0\right) \biggr) +\left(
\Phi_{i},\mathcal{L}^{*} \mathcal{Q}_{i} \Psi _{i}^{(2)}\right) .
\label{c.3}
\end{eqnarray}
The eigenvalue $-\zeta_{0}^{*}/2$ is $d$-fold degenerated and the
convenient choice for the lowest order eigenfunctions has been
discussed in Sec. \ref{s4}. The formal expression for the second
order eigenfunctions $\mathcal{Q}_{i} \Psi_{i}^{(2)}$ is given in
Eq.\ (\ref{5.11a}). For $i\neq 2$, the second term on the right
hand side of Eq.\  (\ref{c.3}) vanishes since $\Phi _{i}$ is an
eigenvector in those cases and, therefore,
\begin{equation}
\left( \Phi_{i},\mathcal{L}^{*} \mathcal{Q}_{i} \Psi_{i}^{(2)}
\right) = \left( \mathcal{L}^{* \dagger} \Phi_{i}, \mathcal{Q}_{i}
\Psi_{i}^{(2)} \right) \propto \left( \Phi_{i}, \mathcal{Q}_{i}
\Psi_{i}^{(2)} \right) =0. \label{c.3a}
\end{equation}
Then, Eq.\ (\ref{c.3}) can be rewritten as
\begin{equation}
\lambda _{i}^{(2)}=\biggl( \widehat{\bm k} \cdot {\bm v}^{\ast
}\Phi _{i},\left\{ \mathcal{Q}_{i} \left[ \mathcal{L}^{\ast
}-\lambda _{i} (0) \right] \right\}^{-1}\widehat{\bm k}\cdot {\bm
v}^{\ast }\Psi _{i}(0) \biggr) +\delta _{i,2}\frac{1}{d}\left(
\mathcal{ L}^{\ast \dagger }v^{\ast 2},\mathcal{Q}_{2}\Psi
_{2}^{(2)}\right), \label{c.4}
\end{equation}
where it has been used that
\begin{equation}
\mathcal{P}_{i}\widehat{\bm k} \cdot {\bm v}^{\ast }\Psi _{i}(0)=
\Psi _{i}(0)\biggl( \Phi _{i},\widehat{\bm k} \cdot {\bm
v}^{\ast}\Psi _{i}(0) \biggr)=0. \label{c.5}
\end{equation}
This follows since $\Phi _{i}$ and $\Psi _{i}(0)$ have the same
parity with respect to reflections of ${\bm v}^{\ast }$.

For the first eigenvalue it is
\begin{eqnarray}
\lambda _{1}^{(2)} &=&\biggl( \widehat{\bm k} \cdot {\bm v}^{\ast
}, \left( \mathcal{Q}_{1}\mathcal{L}^{\ast}\right)^{-1}
\widehat{\bm k}\cdot {\bm v}^{\ast }\Psi _{i}(0) \biggr) = \biggl(
\widehat{\bm k} \cdot {\bm v}^{*}, \mathcal{L}^{*-1} \widehat{\bm
k} \cdot {\bm v}^{*} \Psi_{1}(0) \biggr) \nonumber \\
&=& \biggl( \mathcal{L}^{\ast \dagger -1} \widehat{\bm k} \cdot
{\bm v}^{\ast },\widehat{\bm k} \cdot {\bm v}^{\ast }\Psi _{1} (0)
\biggr) =-\frac{2}{\zeta _{0}^{\ast }} \biggl( ( \widehat{\bm k}
\cdot {\bm v}^{\ast } ) ^{2},\Psi _{1} (0) \biggr) =
\frac{1}{\zeta _{0}^{\ast }}\, . \label{c.6}
\end{eqnarray}
In the first transformation, the property $\mathcal{P}_{1}
\mathcal{L}^{*} X=0$, for arbitrary $X({\bm v}^{*})$, has been
employed. Next, consider the eigenvalue associated to the
longitudinal component of ${\bm \Psi}_{3}(0)$ that we will denote
by $\lambda _{\parallel}^{(2)}$,
\begin{eqnarray}
\lambda _{\parallel}^{(2)} &=&\biggl( \left( \widehat{\bm k}\cdot
{\bm v} ^{\ast }\right) ^{2}, \left[\mathcal{Q}_{3}\left(
\mathcal{L}^{\ast }+\frac{\zeta^{*}_{0}}{2} \right) \right]^{-1}
\widehat{\bm k}\cdot {\bm v}^{\ast } \widehat{\bm k} \cdot
\frac{\partial \ln \phi}{\partial {\bm v}^{*}} \biggr)  \nonumber \\
&=& \biggl( v^{*2}_{i}, \left[ \mathcal{Q}_{3} \left(
\mathcal{L}^{*}+\frac{\zeta_{0}^{*}}{2} \right) \right]^{-1}
G^{*}_{ii} \biggr) \nonumber \\
&& + \biggl( v^{*2}_{i}, \left[ \mathcal{Q}_{3} \left(
\mathcal{L}^{*}+\frac{\zeta^{*}_{0}}{2} \right) \right]^{-1}
\left[ \Psi_{1}(0)+\frac{d+1}{d}\, \Psi_{2}(0) \right] \biggr)
\nonumber \\
&=& \biggl( v^{*2}_{i}, \left(
\mathcal{L}^{*}+\frac{\zeta_{0}^{*}}{2} \right)^{-1} G^{*}_{ii}
\biggr)+\frac{2}{\zeta_{0}^{*}} \biggl( v_{i}^{*2}, \Psi_{1}(0)
\biggr) \nonumber \\
&& + \frac{d-1}{d}\, \frac{1}{\zeta_{0}^{*}} \biggl( v_{i}^{*2},
\Psi_{2}(0) \biggr), \label{c.7}
\end{eqnarray}
where it has been taken into account that $ \mathcal{P}_{3} \left(
\mathcal{L}^{*}+\zeta_{0}^{*}/2 \right)$=0. Using Eq. (\ref{a.18})
in the first term on the right hand side of the above equation and
evaluating explicitly the other two ones one gets
\begin{equation}
\lambda_{\parallel}^{0}= \left( v_{i}^{*2}, C^{*}_{ii} \right)
+\frac{1}{d \zeta_{0}^{*}}= \frac{2(d-1)}{d}\, \eta^{*}
+\frac{1}{d \zeta_{0}^{*}}. \label{c.8}
\end{equation}
The last equality follows from the fact that the expression of
$\eta^{*}$ given in Eq.\ (\ref{a.15}) is equivalent to
\begin{equation}
\left( D^{*}_{ij},C^{*}_{ij} \right) =\eta^{*} \left( 1
+\frac{d-2}{d}\, \delta_{i,j} \right), \label{c.9}
\end{equation}
because of the symmetry of the tensors ${\sf D}$ and ${\sf C}$.
The calculation of the shear modes eigenvalues
$\lambda_{\perp}^{(2)}$ is straightforward:
\begin{eqnarray}
\lambda_{\perp}^{(2)}& =&  \biggl( v^{*}_{\parallel} v^{*}_{\perp
i}\, , \left( \mathcal{L}^{*}+\frac{\zeta_{0}^{*}}{2} \right)^{-1}
v^{*}_{\parallel} \Psi_{3,\perp i} \biggr) \nonumber \\
&=& \biggl( D^{*}_{\parallel,\perp i}\, , \left(
\mathcal{L}^{*}+\frac{\zeta_{0}^{*}}{2} \right)^{-1}
G^{*}_{\parallel,\perp i} \biggr) \nonumber \\
& = & \left( D_{\parallel,\perp i}^{*}\, , C^{*}_{\parallel,\perp
i} \right) = \eta^{*}. \label{c.10}
\end{eqnarray}
Finally, the evaluation of $\lambda_{2}^{(2)}$ is somewhat more
complicated. The contributions from each of the two terms in Eq.
(\ref{c.4}) will be computed separately. The first one is given by
\begin{equation}
\biggl( \widehat{\bm k} \cdot {\bm v}^{\ast }\Phi _{2},  \left[
\mathcal{Q}_{2} \left( \mathcal{L}^{\ast } -
\frac{\zeta_{0}^{*}}{2} \right) \right]^{-1}\widehat{\bm k}\cdot
{\bm v}^{\ast }\Psi _{2}(0) \biggr)
 = \biggl( v^{*}_{i} \Phi_{2},\left( \mathcal{L} -
\frac{\zeta_{0}^{*}}{2} \right)^{-1} v^{*}_{i} \Psi_{2}(0)
\biggr), \label{c.11}
\end{equation}
that is equivalent to
\begin{eqnarray}
& & \frac{1}{d}  \biggl( S^{*}_{i},\left( \mathcal{L} -
\frac{\zeta_{0}^{*}}{2} \right)^{-1} v^{*}_{i} \Psi_{2}(0) \biggr)
+ \frac{d-1}{d} \biggl( v^{*}_{i}, \left( \mathcal{L} -
\frac{\zeta_{0}^{*}}{2} \right)^{-1} v^{*}_{i} \Psi_{2}(0) \biggr)
\nonumber \\
&& \quad \quad \quad = \frac{1}{d}\, \left( S^{*}_{i},
\mathcal{A}_{i}^{*} \right)-\frac{2}{d \zeta_{0}^{*}} \left(
S^{*}_{i}, \Psi_{3,i}(0) \right) - \frac{d+1}{d}\,
\frac{1}{\zeta_{0}^{*}} \biggl( v^{*}_{i},v^{*}_{i} \Psi_{2}(0)
\biggr) \nonumber \\
&& \quad \quad \quad = \kappa^{*}-\frac{d+1}{d \zeta^{*}_{0}}\, .
\label{c.12}
\end{eqnarray}
The analysis of the second term on the right hand side of Eq.\
(\ref{c.4}) is carried out in an analogous way,
\begin{equation}
\frac{1}{d} \left( \mathcal{L}^{*} v^{*2},\mathcal{Q}_{2}
\Psi_{2}^{(2)} \right)= -\frac{\zeta_{0}^{*}}{d} \left(
v^{*2},\mathcal{Q}_{2} \Psi_{2}^{(2)} \right)+\frac{1}{d}
\left(a,\mathcal{Q}_{2} \Psi_{2}^{(2)} \right). \label{c.13}
\end{equation}
It is
\begin{eqnarray}
-\frac{\zeta_{0}^{*}}{d} \left( v^{*2},\mathcal{Q}_{2}^{2}
\Psi_{2}^{(2)}\right) & = & \frac{\zeta_{0}^{*}}{2}
\left(\Phi_{1},\mathcal{Q}_{2} \Psi_{2}^{(2)} \right) \nonumber \\
& = & - \frac{1}{\zeta_{0}^{*}} \left( \Phi_{3,i},v^{*}_{i}
\Psi_{2}(0) \right) = - \frac{1}{\zeta_{0}^{*}}. \label{c.14}
\end{eqnarray}
In the second equality above, the explicit expression of
$\Psi_{2}^{(2)}$ given in Eq.\ (\ref{5.11a}) as well as the
properties of $\mathcal{L}^{*}$ have been used. The second term on
the right hand side of Eq.\ (\ref{c.13}) can be rewritten as
\begin{equation}
\frac{1}{d} \left(a,\mathcal{Q}_{2} \psi_{2}^{(2)} \right)=
-\frac{1}{d} \left( a,\mathcal{M}^{*}_{1} \right), \label{c.15}
\end{equation}
where $\mathcal{M}^{*}_{1}$ is defined in Eq.\ (\ref{a.40}).
Substitution of Eqs.\ (\ref{c.14}) and (\ref{c.15}) into Eq.\
(\ref{c.13}) yields
\begin{equation}
\frac{1}{d} \left( \mathcal{L}^{*} v^{*2},\mathcal{Q}_{2}
\Psi_{2}^{(2)} \right)= -\frac{1}{\zeta_{0}^{*}}-\frac{1}{d}
\left( a,\mathcal{M}^{*}_{1} \right)=-\zeta^{*}_{1}, \label{c.16}
\end{equation}
because of Eq.\ (\ref{a.39}). Then, use of Eqs.\ (\ref{c.12}) and
(\ref{c.16}) into Eq.\ (\ref{c.4}) gives the final expression for
$\lambda_{2}^{(2)}$,
\begin{equation}
\lambda_{2}^{(2)}=\kappa^{*}-\zeta_{1}^{*}-\frac{d+1}{d
\zeta_{0}^{*}}. \label{c.16a}
\end{equation}
Comparison of the results for the second order eigenvalues
obtained in this Appendix with those reported in Sec.  \ref{s2}
shows that both agree, with the transport coefficients given by
the same expressions in both cases.

\section{Linearization of the Model Kinetic Equation}
\label{ap4}

Solutions to the model kinetic equation (\ref{6.1}) are sought of
the form (\ref{3.4}) and, consistently, $g=g_{hcs}+\delta g$ and
$\nu =\nu_{hcs}+\delta \nu$, retaining only terms linear in
$\Delta $,
\begin{equation}
\left( \frac{\partial }{\partial t}+\mathbf{v}\cdot \nabla \right)
(f_{hcs}\Delta) =-\delta \nu \left( f_{hcs}-g_{hcs}\right) -\nu
_{hcs}\left( f_{hcs}\Delta -\delta g\right) .  \label{d.1}
\end{equation}
Use has been made of the HCS model equation
\begin{equation}
\frac{\partial f_{hcs}}{\partial t} = \frac{\zeta_{hcs}}{2}
\frac{\partial}{\partial {\bm v}} \cdot  \left( {\bm v}
f_{hcs}\right) =-\nu _{hcs}\left( f_{hcs}-g_{hcs}\right)
.\hspace{0.2in} \label{d.2}
\end{equation}
Next, from the choice of $\nu$ as corresponding to hard sphere
behavior,
\begin{equation}
\delta \nu =\nu _{hcs}\left( \frac{\delta n}{n_{hcs}}+\frac{2
\delta T}{T_{hcs}}\right) . \label{d.3}
\end{equation}
Similarly, the definition of $g$ in Eq.\ (\ref{6.2}) gives
\begin{equation}
\delta g =\frac{g_{hcs}}{n_{hcs}}\, \delta n + \frac{\partial
g_{hcs}}{ \partial T_{hcs}}\, \delta T- \frac{\partial
g_{hcs}}{\partial {\bm v}}\, \cdot {\bm u}. \label{d.4}
\end{equation}
The linear model kinetic equation (\ref{d.1}) then becomes
\begin{eqnarray}
\left( \frac{\partial }{\partial t}+\mathbf{v}\cdot \nabla \right)
( f_{hcs}\Delta)  &=&-\left( \frac{\delta n}{n_{hcs}}+\frac{\delta
T}{2T_{hcs}}\right) \nu _{hcs}\left( f_{hcs}-g_{hcs}\right)  \nonumber \\
&&-\nu _{hcs}\left( f_{hcs}\Delta -\frac{g_{hcs}}{n_{hcs}}\,
\delta n -\frac{\partial g_{hcs}}{\partial T_{hcs}}\, \delta T +
\frac{\partial g_{hcs}}{\partial {\bm v}} \cdot {\bm u} \right).
\label{d.5}
\end{eqnarray}
It is convenient to eliminate $g_{hcs}$ in this result using the
HCS equation (\ref{d.2}) to get
\begin{eqnarray}
\left( \frac{\partial }{\partial t}+\mathbf{v}\cdot \nabla \right)
(f_{hcs}\Delta) &=&-\nu _{hcs}f_{hcs}\Delta +\left( \frac{\delta
n}{n_{hcs}}+ \frac{\delta T}{2T_{hcs}}\right)
\frac{\zeta_{hcs}}{2} \frac{\partial}{\partial {\bm v}}
\cdot \left( {\bm v} f_{hcs}\right)  \nonumber \\
&&+\frac{\delta n}{n_{hcs}}\left[ \nu
_{hcs}f_{hcs}+\frac{\zeta_{hcs}}{2} \frac{\partial}{\partial {\bm
v}} \cdot \left( {\bm v}f_{hcs}\right) \right]
\nonumber \\
&&+\delta T\left[ \nu _{hcs}\frac{\partial f_{hcs}}{
\partial T_{hcs}}+\frac{\zeta _{hcs}}{2}
\frac{\partial}{\partial {\bm v}} \cdot \left( {\bm
v}\frac{\partial f_{hcs}}{\partial T_{hcs}}\right) \right]
\nonumber \\
&&-{\bm u}\cdot \frac{\partial}{\partial {\bm v}} \left[ \nu
_{hcs}f_{hcs}+\frac{\zeta _{hcs}}{2} \frac{\partial}{\partial {\bm
v}} \cdot \left( {\bm v}f_{hcs}\right) \right].  \label{d.6}
\end{eqnarray}
The right hand side of this expression can be put in a more
transparent form by noting that
\begin{equation}
\left\{ \left( \Delta ,\Phi _{i}\right) \right\} =\left\{
\frac{\delta n}{ n_{hcs}}, \frac{\delta T}{2T_{hcs}}+\frac{\delta
n}{n_{hcs}},\frac{u_{\parallel}}{v_{hcs}},\frac{{\bm
u}_{\perp}}{v_{hcs}} \right\} . \label{d.7}
\end{equation}
where the scalar product $(X,Y)$ is defined in Eq.\ (\ref{3.11})
and $\left\{ \Phi _{i}\right\} $ are the biorthogonal set given in
Eq.\  (\ref{4.14}). Moreover, the relation
\begin{equation}
T_{hcs}\, \frac{\partial \ln f_{hcs}}{\partial T_{hcs}}=\frac{1}{2
\phi } \frac{\partial}{\partial {\bm v}} \cdot \left( {\bm v}\phi
\right), \label{d.8}
\end{equation}
allows to write the dependence on $f_{hcs}$ on the right hand side
in terms of the lowest order hydrodynamic eigenfunctions $\{
\Psi_{i}(0) \}$ of the linearized Boltzmann operator given in Eq.\
(\ref{4.10}). Then, Eq.\ (\ref{d.6}) can be rewritten as
\begin{eqnarray}
\left( \frac{\partial }{\partial t}+ {\bm v} \cdot \nabla \right)
( f_{hcs}\Delta)  &=&-\nu _{hcs}f_{hcs}\left[ \Delta -\sum_{i}
\Psi_{i}(0) \left(\Phi _{i}, \Delta \right) \right] \nonumber
\\  & & + \frac{\zeta _{hcs}}{2} \frac{\partial}{\partial {\bm v}} \cdot
\left[ f_{hcs}{\bm v} \sum_{i} \Psi_{i}(0) \left(\Phi _{i}, \Delta
\right) \right] \nonumber \\
&&-f_{hcs}\frac{v_{hcs}(t)}{\ell} \sum_{i}\lambda _{i}(0)\Psi
_{i}(0)\left( \Phi _{i},\Delta \right) .  \label{d.10}
\end{eqnarray}
Here $\left\{ \lambda _{i}(0)\right\} $ are the eigenvalues of Eq.
(\ref{4.9}).

Finally, introducing the dimensionless variables of Secs. \ref{s2}
and \ref{s3}, and the single Fourier component of Eq.\ (\ref{3.7})
gives the linear kinetic equation for the model
\begin{equation}
\left( \partial _{s}+i\mathbf{k\cdot v}^{\ast
}\mathbf{+}\mathcal{L}^{\ast }_{m}\right) \tilde{\Delta} ({\bm
k},{\bm v}^{\ast},s)=0, \label{d.12}
\end{equation}
with
\begin{equation}
\mathcal{L}^{\ast }_{m}=\sum_{i}\lambda _{i}(0)\mathcal{P}_{i}+\nu
_{0}^{\ast } \mathcal{Q}+\frac{\zeta _{0}^{\ast }}{2}\phi ^{-1}
\frac{\partial}{\partial {\bm v}^{*}} \cdot {\bm v}^{\ast }\phi
\mathcal{Q}. \label{d.13}
\end{equation}
The projection operators $\mathcal{P}_{i}$ are defined in Eq.\
(\ref{5.6}), and
\begin{equation}
\mathcal{P} \equiv \sum_{i}\mathcal{P}_{i},\quad \mathcal{Q}
\equiv 1-\mathcal{P}. \label{d.15}
\end{equation}
It is directly verified that
\begin{equation}
\mathcal{P} \phi^{-1} \frac{\partial}{\partial {\bm v}^{*}} \cdot
{\bm v}^{*} \phi \mathcal{Q} =0, \label{d.15a}
\end{equation}
so that $\mathcal{L}^{\ast }_{m}$ can be written
\begin{equation}
\mathcal{L}^{\ast }_{m}=\sum_{i}\lambda
_{i}(0)\mathcal{P}_{i}+\mathcal{Q}\left[ \nu _{0}^{\ast
}+\frac{\zeta _{0}^{\ast }}{2} \phi ^{-1} \frac{\partial}{\partial
{\bm v}^{*}} \cdot {\bm v}^{\ast }\phi \right] \mathcal{Q}.
\label{d.16}
\end{equation}
This is the expression for the generator of the linear dynamics
used in main the text.

\end{document}